\documentclass[english,aps,manuscript,A4, english, 12pt]{revtex4}
\usepackage[T1]{fontenc}
\usepackage[latin9]{inputenc}
\usepackage{color}
\usepackage{babel}
\usepackage{amsmath}
\usepackage{graphicx}
\usepackage[unicode=true,pdfusetitle,
 bookmarks=true,bookmarksnumbered=false,bookmarksopen=true,bookmarksopenlevel=1,
 breaklinks=false,pdfborder={0 0 0},pdfborderstyle={},backref=false,colorlinks=true]
 {hyperref}
\hypersetup{
 linkcolor=blue, citecolor=cyan}

\makeatletter

\providecommand{\tabularnewline}{\\}

\@ifundefined{textcolor}{}
{%
 \definecolor{BLACK}{gray}{0}
 \definecolor{WHITE}{gray}{1}
 \definecolor{RED}{rgb}{1,0,0}
 \definecolor{GREEN}{rgb}{0,1,0}
 \definecolor{BLUE}{rgb}{0,0,1}
 \definecolor{CYAN}{cmyk}{1,0,0,0}
 \definecolor{MAGENTA}{cmyk}{0,1,0,0}
 \definecolor{YELLOW}{cmyk}{0,0,1,0}
}

\makeatother

\begin{document}
\title{Lepton pair photoproduction in peripheral relativistic heavy-ion collisions}
\author{Ren-jie Wang}
\email{wrjn@mail.ustc.edu.cn}

\affiliation{Department of Modern Physics, University of Science and Technology
of China, Hefei, Anhui 230026, China}
\author{Shuo Lin}
\email{linshuo@mail.ustc.edu.cn}

\affiliation{Department of Modern Physics, University of Science and Technology
of China, Hefei, Anhui 230026, China}
\author{Shi Pu}
\email{shipu@ustc.edu.cn}

\affiliation{Department of Modern Physics, University of Science and Technology
of China, Hefei, Anhui 230026, China}
\author{Yi-fei Zhang}
\email{ephy@ustc.edu.cn}

\affiliation{Department of Modern Physics, University of Science and Technology
of China, Hefei, Anhui 230026, China}
\author{Qun Wang}
\email{qunwang@ustc.edu.cn}

\affiliation{Department of Modern Physics, University of Science and Technology
of China, Hefei, Anhui 230026, China}
\begin{abstract}
We study the lepton pair photoproduction in peripheral heavy-ion collisions
based on the formalism in our previous work \citep{Wang:2021kxm}.
We present the numerical results for the distributions of the transverse
momentum, azimuthal angle and invariant mass for $e^{+}e^{-}$ and
$\mu^{+}\mu^{-}$ pairs as functions of the impact parameter and other
kinematic variables in Au+Au collisions. Our calculation incorporates
the information on the transverse momentum and polarization of photons
which is essential to describe the experimental data. We observe a
broadening effect in the transverse momentum for lepton pairs with
and without smear effects. We also observe a significant enhancement
in the distribution of $\cos(2\varphi)$ for{\normalsize{} }$\mu^{+}\mu^{-}$
pairs. Our results provide a baseline for future studies of other
higher order corrections beyond Born approximation and medium effects
in the lepton pair production.
\end{abstract}
\maketitle

\section{Introduction \label{sec:Introduction}}

Extremely strong electromagnetic fields are generated in relativistic
heavy-ion collisions \citep{Kharzeev:2007jp,Skokov:2009qp,Bzdak:2011yy,Voronyuk:2011jd,Deng:2012pc,Roy:2015coa,Li:2016tel,Inghirami:2016iru,Roy:2015kma,Pu:2016ayh,Pu:2016bxy,Pu:2016rdq,Siddique:2019gqh,Wang:2020qpx}
and induce many novel quantum transport phenomena such as the chiral
magnetic effect, the chiral separation effect \citep{Vilenkin:1980fu,Kharzeev:2007jp,Fukushima:2008xe},
the chiral electric separation effect \citep{Huang:2013iia,Pu:2014cwa,Jiang:2014ura},
and other nonlinear chiral effects \citep{Chen:2016xtg,Pu:2014fva,Hidaka:2017auj,Ebihara:2017suq,Chen:2013dca}.
For more discussions of these chiral effects, we refer readers to
recent reviews \citep{Kharzeev:2015znc,Liao:2014ava,Miransky:2015ava,Huang:2015oca,Fukushima:2018grm,Bzdak:2019pkr,Zhao:2019hta,Gao:2020vbh,Hidaka:2022dmn}
and references therein. Meanwhile, some quantum electrodynamics (QED)
effects have also been widely discussed, such as the light-by-light
scattering \citep{Aaboud:2017bwk}, matter generation directly from
photons \citep{Adam:2019mby}, the vacuum birefringence \citep{Hattori:2012je,Hattori:2012ny,Hattori:2020htm,Adam:2019mby}
and the Schwinger mechanism \citep{Schwinger:1951nm,Copinger:2018ftr,Copinger:2020nyx, Copinger:2022jgg}. 

According to the equivalent photon approximation (EPA) or the Weizsaecker-Williams
method \citep{vonWeizsacker:1934nji,Williams:1934ad}, strong electromagnetic
fields generated by a fast moving nucleus can be treated as the quasi-real
photons. Recently, the lepton pair production through the fusion of
two quasi-real photons, the named Breit-Wheeler process \citep{Breit:1934zz},
has been measured in several experiments of relativistic heavy-ion
collisions. The STAR Collaboration at Relativistic heavy-ion Collider
(RHIC) has measured the Breit-Wheeler process in ultra-peripheral
collisions (UPC) \citep{Adams:2004rz,Adam:2019mby} and peripheral
collisions \citep{Adam:2018tdm}. A significant enhancement of the
lepton pair production at low transverse momenta of dileptons ($P_{T}<0.15$
$\textrm{GeV}$) in peripheral collisions has been measured in the
STAR experiment \citep{Adam:2018tdm} in comparison with the hadronic
cocktails. The azimuthal asymmetry of the lepton pair originated from
the linear polarization of incoming photons has also been observed
by STAR \citep{Adam:2019mby} in connection with the vacuum birefringence
phenomena \citep{Heisenberg:1936nmg}. Furthermore the broadening
effects are seen in peripheral collisions by STAR \citep{Adam:2018tdm}
at RHIC, ATLAS \citep{ATLAS:2018pfw} and CMS at the Large Hadron
Collider \citep{CMS:2020skx}.

We note that the EPA fails to describe experimental data \citep{Klein:2016yzr,Adam:2018tdm}
due to missing the essential information of the transverse momentum
and polarization of photons. To better describe the data, several
theoretical methods have been proposed. These include a generalized
EPA (gEPA) or a QED-based method in the background field approach
\citep{Vidovic:1992ik,Hencken:1994my,Hencken:2004td,Zha:2018tlq,Zha:2018ywo,Brandenburg:2020ozx,Brandenburg:2021lnj},
the transverse momentum dependent (TMD) parton distribution functions
\citep{Li:2019yzy,Li:2019sin}, the factorization formalism with the
photon Wigner functions \citep{Klein:2020jom,Xiao:2020ddm}, and the
wave-packet method based on the classical field approximation in QED
\citep{Wang:2021kxm}.

Although these theoretical methods can account for most experimental
data, the broadening effect for the transverse momentum of the dilepton
pair in peripheral collisions has not been fully understood. There
are two extra sources to this effect, the higher order correction
beyond the Born approximation and the medium effect. The Sudakov factor
as a sum over the soft photon emission is a typical higher order correction
beyond the Born approximation \citep{Klein:2018fmp,Klein:2020jom,Li:2019sin,Li:2019yzy,Hatta:2021jcd}.
There are other effects from the medium, such as the effect from the
Lorentz force \citep{Adam:2018tdm}, multiple scatterings of lepton
pairs in the medium \citep{Klein:2020jom,Klein:2018fmp,ATLAS:2018pfw},
that may also lead to the broadening of transverse momenta.

Meanwhile, although the azimuthal angle distribution of $e^{+}e^{-}$
pairs has been predicted and confirmed in experiments \citep{Li:2019yzy,Li:2019sin},
a systematic study of other lepton species such as $\mu^{+}\mu^{-}$
pairs is still needed for future experiments \citep{Zhou:2022gbh}.

In this paper, we extend our previous study of the UPC \citep{Wang:2021kxm}
to peripheral collisions. In our approach, we will calculate the broadening
effect and the azimuthal angle distribution for $e^{+}e^{-}$ and
$\mu^{+}\mu^{-}$ pairs for different centralities in Au+Au collisions
at $\sqrt{s_{NN}}=200$ GeV. We expect to observe the enhancement
of the $\cos(2\varphi)$ modulation in $\mu^{+}\mu^{-}$ pairs due
to the mass effect as proposed in Ref. \citep{Li:2019yzy,Li:2019sin}.
Similar azimuthal asymmetry have also found in vector meson photoproduction \citep{Xing:2020hwh,Hagiwara:2020juc, STAR:2022wfe,Brandenburg:2022jgr}.
For the broadening effect, we will consider a smearing of the lepton
momentum from the finite resolution of the measured momentum and multiple
scatterings in the detector as well as from the bremsstrahlung radiation
of electrons. All these effects have been usually considered in experiments.
In this work, as a first attempt, we add the smearing effect extracted
from the measurement of $J/\psi\rightarrow l\overline{l}$ by STAR
\citep{STAR:2005gfr} to the transverse momentum distribution.

The paper is organized as follows. In Sec. \ref{sec: Cross sections},
we give a brief overview on our previous work \citep{Wang:2021kxm}.
After introducing parameters used in our calculation in Sec. \ref{subsec:Parameters},
we compute the transverse momentum broadening effect in Sec. \ref{subsec:Transverse-momentum-distribution},
the azimuthal angle distributions for $e^{+}e^{-}$ and $\mu^{+}\mu^{-}$
pairs in Sec. \ref{subsec:Azimuthal-angle-distribution}, and the
invariant mass distribution in Sec. \ref{subsec:Invariant-mass-distribution}.
We make a summary of our results in Sec. \ref{sec:Summary}. Throughout
this paper, we use the sign convention for the metric tensor $g_{\mu\nu}=\textrm{diag}\{+,-,-,-\}$.


\section{Cross sections for lepton pair photoproduction \label{sec: Cross sections}}

In this section, we give a brief overview on differential cross sections
for lepton pair photoproduction at Born level derived in our previous
work \citep{Wang:2021kxm}.

We consider head-on collisions of two identical nuclei $A_{1}$ and
$A_{2}$ moving in $\pm z$ direction in which a pair of leptons ($l$
and $\overline{l}$) are generated accompanied by other particles
$X_{1},\cdots,X_{f}$. The process can be expressed by 
\begin{equation}
A_{1}(P_{1})+A_{2}(P_{2})\rightarrow l(k_{1})+\overline{l}(k_{2})+\sum_{f}X_{f}(K_{f}),
\end{equation}
where the particles' four-momenta are given in parentheses. The on-shell
momenta of two nuclei are denoted as $P_{i}^{\mu}=(E_{Pi},\mathbf{P}_{i})=Mu_{i}^{\mu}$
for $i=1,2$, where $E_{Pi}=\sqrt{\mathbf{P}_{i}^{2}+M^{2}}$, $u_{1,2}^{\mu}=\gamma\left(1,0,0,\pm v\right)$
with $\gamma=1/\sqrt{1-v^{2}}$, and $M$ are the energies, four-velocities
and mass of two nuclei. The three-momenta of two nuclei are $\mathbf{P}_{1}=(0,0,P_{1}^{z})$
and $\mathbf{P}_{2}=(0,0,-P_{1}^{z})$ in the center of mass frame
of the collision. We consider the photon fusion as the sub-process
of the collision
\begin{equation}
\gamma(p_{1})+\gamma(p_{2})\rightarrow l(k_{1})+\overline{l}(k_{2}).\label{eq: Subprocess}
\end{equation}
Note that each photon does not need to be emitted from the center
of a nucleus.


In our previous work \citep{Wang:2021kxm}, we empoly the narrow wave-packet
to describe two colliding nuclei. The differential cross section for
the photoproduction of lepton pairs can be put into a compact form
\begin{eqnarray}
\frac{d\sigma}{d^{3}k_{1}d^{3}k_{2}} & = & \frac{1}{32(2\pi)^{6}}\frac{1}{vE_{P1}E_{P2}}\frac{1}{E_{k1}E_{k2}}\int d^{2}\mathbf{b}_{T}d^{2}\mathbf{b}_{1T}d^{2}\mathbf{b}_{2T}\int d^{4}p_{1}d^{4}p_{2}\nonumber \\
 &  & \times\delta^{(2)}\left(\mathbf{b}_{T}-\mathbf{b}_{1T}+\mathbf{b}_{2T}\right)(2\pi)^{4}\delta^{(4)}\left(p_{1}+p_{2}-k_{1}-k_{2}\right)\nonumber \\
 &  & \times\mathcal{S}_{\sigma\mu}(p_{1},\mathbf{b}_{1T})\mathcal{S}_{\rho\nu}(p_{2},\mathbf{b}_{2T})\nonumber \\
 &  & \times\sum_{\textrm{spin of }l,\overline{l}}L^{\mu\nu}(p_{1},p_{2};k_{1},k_{2})L^{\sigma\rho*}(p_{1}^{\prime},p_{2}^{\prime};k_{1},k_{2}),\label{eq: QED cross section}
\end{eqnarray}
where we have introduced the transverse position $\mathbf{b}_{iT}$
of the photon emission in the nucleus $A_{i}$, $\mathbf{b}_{T}=\mathbf{b}_{1T}-\mathbf{b}_{2T}$
is the impact parameter of the colliding nuclei, the lepton tensor
$L^{\mu\nu}$ at the tree or Born level is given by
\begin{eqnarray}
L^{\mu\nu}(p_{1},p_{2};k_{1},k_{2}) & = & -ie^{2}\overline{u}(k_{1})\left[\gamma^{\mu}\frac{\gamma\cdot(k_{1}-p_{1})+m}{(k_{1}-p_{1})^{2}-m^{2}+i\varepsilon}\gamma^{\nu}\right.\nonumber \\
 &  & \left.+\gamma^{\nu}\frac{\gamma\cdot(p_{1}-k_{2})+m}{(p_{1}-k_{2})^{2}-m^{2}+i\varepsilon}\gamma^{\mu}\right]v(k_{2}),\label{eq: Lepton parts}
\end{eqnarray}
and $\mathcal{S}_{\sigma\mu}(p_{i},\mathbf{b}_{iT})$ denotes the
Born-level Wigner function for photons \citep{Klein:2020jom,Xiao:2020ddm}
\begin{equation}
\mathcal{S}_{\sigma\mu}(p_{i},\mathbf{b}_{iT})\equiv\int\frac{d^{2}\boldsymbol{\Delta}_{iT}}{(2\pi)^{2}}\int\frac{d^{4}y_{i}}{(2\pi)^{4}}e^{ip_{i}\cdot y_{i}}\left\langle P_{i}^{\prime}\right|A_{\sigma}^{\dagger}(0)A_{\mu}\left(y_{i}\right)\left|P_{i}\right\rangle e^{-i\mathbf{b}_{iT}\cdot\boldsymbol{\Delta}_{iT}}.\label{eq: Wigner functions}
\end{equation}
which encodes the information of transverse phase space for photons
and leads to the transverse momentum broadening. In Eq. (\ref{eq: Wigner functions}),
$P_{i}^{\prime}\equiv(E_{Pi},\mathbf{P}_{iT}+\boldsymbol{\Delta}_{iT},P_{iz})$
is connected to $p_{i}^{\prime}$ in Eq. (\ref{eq: QED cross section})
by $p_{i}+p_{i}^{\prime}=P_{i}+P_{i}^{\prime}$, and the transverse
momentum dependent (TMD) distribution function for photons at Born
level \citep{Li:2019sin,Li:2019yzy} can be obtained from $\mathcal{S}_{\sigma\mu}(p_{i},\mathbf{b}_{iT})$
after integration over $\mathbf{b}_{iT}$. Equation (\ref{eq: QED cross section})
is very close to the differential cross section in the framework of
TMD factorization \citep{Xiao:2020ddm,Li:2019sin,Li:2019yzy,Klein:2018fmp,Klein:2020jom}.
For other  discussions on the 
geometric relations for ions or protons can also be found in Ref. \citep{Baur:2001jj,Baltz:2009jk,Klein:2016yzr,Klein:2018cjh,Zha:2018ywo,Klein:2020jom,Klusek-Gawenda:2020eja,Godunov:2021pdz}.

We adopt the classical field approximation for $\mathcal{S}_{\sigma\mu}(p_{i},\mathbf{b}_{iT})$,
then the cross section (\ref{eq: QED cross section}) can be put into
the form
\begin{eqnarray}
\sigma & = & \frac{Z^{4}e^{4}}{2\gamma^{4}v^{3}}\int d^{2}\mathbf{b}_{T}d^{2}\mathbf{b}_{1T}d^{2}\mathbf{b}_{2T}\int\frac{d\omega_{1}d^{2}\mathbf{p}_{1T}}{(2\pi)^{3}}\frac{d\omega_{2}d^{2}\mathbf{p}_{2T}}{(2\pi)^{3}}\nonumber \\
 &  & \times\int\frac{d^{2}\mathbf{p}_{1T}^{\prime}}{(2\pi)^{2}}e^{-i\mathbf{b}_{1T}\cdot(\mathbf{p}_{1T}^{\prime}-\mathbf{p}_{1T})}\frac{F^{*}(-\overline{p}_{1}^{\prime2})}{-\overline{p}_{1}^{\prime2}}\frac{F(-\overline{p}_{1}^{2})}{-\overline{p}_{1}^{2}}\nonumber \\
 &  & \times\int\frac{d^{2}\mathbf{p}_{2T}^{\prime}}{(2\pi)^{2}}e^{-i\mathbf{b}_{2T}\cdot(\mathbf{p}_{2T}^{\prime}-\mathbf{p}_{2T})}\frac{F^{*}(-\overline{p}_{2}^{\prime2})}{-\overline{p}_{2}^{\prime2}}\frac{F(-\overline{p}_{2}^{2})}{-\overline{p}_{2}^{2}}\nonumber \\
 &  & \times\int\frac{d^{3}k_{1}}{(2\pi)^{3}2E_{k1}}\frac{d^{3}k_{2}}{(2\pi)^{3}2E_{k2}}\nonumber \\
 &  & \times\sum_{\textrm{spin of }l,\overline{l}}\left[u_{1\mu}u_{2\nu}L^{\mu\nu}(\overline{p}_{1},\overline{p}_{2};k_{1},k_{2})\right]\left[u_{1\sigma}u_{2\rho}L^{\sigma\rho*}(\overline{p}_{1}^{\prime},\overline{p}_{2}^{\prime};k_{1},k_{2})\right]\nonumber \\
 &  & \times(2\pi)^{4}\delta^{(4)}\left(\overline{p}_{1}+\overline{p}_{2}-k_{1}-k_{2}\right)\delta^{(2)}\left(\mathbf{b}_{T}-\mathbf{b}_{1T}+\mathbf{b}_{2T}\right),\label{eq:main-cross-section}
\end{eqnarray}
where $F(p)$ is the charge form factor of the nucleus, i.e. the Fourier
transformation of nuclear charge density, $\overline{p}_{i}$ and
$\overline{p}_{i}^{\prime}$ satisfying $\overline{p}_{i}\cdot u_{i}=\overline{p}_{i}^{\prime}\cdot u_{i}=0$
are photon momenta in the classical field approximation which can
further be written as
\begin{eqnarray}
\overline{p}_{i}^{\mu} & = & \left(\omega_{i},\boldsymbol{p}_{iT},(-1)^{i+1}\frac{\omega_{i}}{v}\right),\quad\overline{p}_{i}^{\prime\mu}=\left(\omega_{i},\boldsymbol{p}_{iT}^{\prime},(-1)^{i+1}\frac{\omega_{i}}{v}\right).
\end{eqnarray}

Equation (\ref{eq:main-cross-section}) incorporates the information
of the position, momentum and polarization of the photons. If we integrate
over $\mathbf{b}_{iT}$, Eq. (\ref{eq:main-cross-section}) reduces
to the differential cross section used in recent studies \citep{Zha:2018tlq,Zha:2018ywo,Brandenburg:2020ozx,Brandenburg:2021lnj}
based on QED models in the classical background field approach \citep{Vidovic:1992ik,Hencken:1994my,Hencken:2004td}.
In the ultra-relativistic limit, Eq. (\ref{eq:main-cross-section})
gives the same expression as EPA \citep{Baltz:2009jk,Klein:2016yzr,Zha:2018ywo}
or gEPA \citep{Zha:2018tlq,Zha:2018ywo,Brandenburg:2020ozx,Brandenburg:2021lnj}.
In the twist expansion of the photon's Wigner function, Eq. (\ref{eq:main-cross-section})
is equivalent to the differential cross section at Born level in the
TMD framework up to the twist-2 \citep{Xiao:2020ddm,Li:2019sin,Li:2019yzy,Klein:2018fmp,Klein:2020jom}.


\section{Numerical results \label{sec:Numerical-results}}

In this section, we present numerical results based on Eq. (\ref{eq:main-cross-section})
for the spectra of the transverse momentum, azimuthal angle and invariant
mass for lepton pairs in Au+Au collisions at 200 GeV. The results
are compared with the experimental data in Refs. \citep{Adam:2018tdm,Adam:2019mby}.

\begin{table}
\caption{Centralities and impact parameters in Au+Au collisions from Ref. \citep{Klein:2018cjh}.
\label{tab: Centralities}}

\setlength{\tabcolsep}{30pt}

\begin{tabular}{c|c}
\hline 
Centralities & Impact parameters\tabularnewline
\hline 
\hline 
UPC RHIC Au+Au & $>2R_{A}$\tabularnewline
\hline 
60-80\% RHIC Au+Au  & $11.4-13.2$ fm\tabularnewline
\hline 
40-60\% RHIC Au+Au & $9.4-11.6$ fm\tabularnewline
\hline 
10-40\% RHIC Au+Au & $4.8-9.4$ fm\tabularnewline
\hline 
\end{tabular}
\end{table}

\subsection{Parameters \label{subsec:Parameters}}

In this subsection, we show the parameters used in the numerical calculation.
The charge form factor for the nucleus is chosen to be 
\begin{equation}
F(p)=\frac{4\pi\rho^{0}}{p^{3}A}[\sin(pR_{A})-pR_{A}\cos(pR_{A})]\frac{1}{a^{2}p^{2}+1},\label{eq:form factor}
\end{equation}
where $a=0.7$ fm, $A$ is the number of nucleons in the nucleus,
$R_{A}=1.2A^{1/3}$ fm is the nucleus radius, and $\rho^{0}=3A/(4\pi R_{A}^{3})$
is the nucleon number density which makes $F(p=0)=1$ \citep{Klein:1999qj}.
The form factor (\ref{eq:form factor}) is very close to the Fourier
transformation of the Woods-Saxson distribution that has been used
in the calculation of photoproduction and other processes \citep{Klein:2016yzr,Klein:1999qj,Li:2019sin,Li:2019yzy}.

When the impact parameter $b_{T}$ is larger than $\sim13.2$ fm corresponding
to $\geq80\%$ centrality, two colliding nuclei may undergo mutual
Coulomb excitation and emit neutrons \citep{Li:2019sin,Brandenburg:2020ozx}.\textcolor{blue}{{}
}To account for the effect of such processes, one needs to multiply
an extra pre-factor $\mathcal{P}^{2}(b_{T})$ to the cross section,
where $\mathcal{P}(b_{T})$ is the probability of emitting neutrons
from an excited nucleus. It can be parametrized as \citep{Bertulani:1987tz},
\begin{equation}
\mathcal{P}(b_{T})=\sum_{N_{\gamma}=1}^{\infty}\frac{1}{N_{\gamma}!}w^{N_{\gamma}}\exp(-w)=1-\exp(-w),\label{eq:P_b_perp}
\end{equation}
where $N_{\gamma}$ denotes the number of photons that can be absorbed
by a nucleus. In our previous work \citep{Wang:2021kxm}, we have
chosen the $w$ from the Giant Dipole Resonance model \citep{Bertulani:1987tz,Baur:1998ay,Hencken:2004td,Baltz:2009jk,Li:2019sin},
$w=5.45\times10^{-5}Z^{3}(A-Z)/[A^{2/3}b_{T}^{2}]$. In the comparison
with the experimental measurement, one needs to consider the case
of emitting multiple neutrons and rewrite the $w$ as \citep{Brandenburg:2020ozx,Baltz:1998ex,Baltz:2002pp,Broz:2019kpl,Baltz:2009jk,Pshenichnov:2001qd}
\begin{equation}
w_{Xn}(b_{T})=\int d\omega n(\omega,b_{T})\sigma_{\gamma+A\rightarrow A^{\prime}+Xn}(\omega),\label{eq:XnXn_w}
\end{equation}
where the $X\geq1$ stands for the number of neutrons emitted by a
nuclei and $n(\omega,b_{T})$ is the photon flux \citep{Jackson:1998nia,Wang:2021kxm,Vidovic:1992ik}
and the photon-nucleus cross section $\sigma_{\gamma+A\rightarrow A^{\prime}+Xn}$
is given by the fixed-target experiments \citep{Veyssiere:1970ztg,Berman:1987zz,Baltz:1998ex}.
One can also approximates $\mathcal{P}(b_{T})\approx w$ \citep{Hencken:2004td}
or $\mathcal{P}(b_{T})\approx w\exp(-w)$ \citep{Bertulani:1987tz,Baltz:1996as,Baur:1998ay,Brandenburg:2020ozx}
for small $w$ or $N_{\gamma}\approx1$ respectively. In the current
work, we have chosen $\mathcal{P}(b_{T})$ in Eq. (\ref{eq:P_b_perp})
with $w$ given by Eq. (\ref{eq:XnXn_w}) to describe the multiple
neutrons emission processes in the experiments, when the impact parameter
is larger than $\sim13.2$ fm. 

Since the differential cross section (\ref{eq:main-cross-section})
invloves a high-dimension integration, we employ the algorithm of
ZMCintegral \citep{Wu_2020cpc_wzpw,Zhang_2020cpc_zw} to handle such
a high-dimension integral. Other applications of ZMCintegral to relativistic
heavy-ion collisions can be found in Ref. \citep{Zhang:2019uor,Zhang:2022lje}.

\begin{figure}[t]
\includegraphics[scale=0.3]{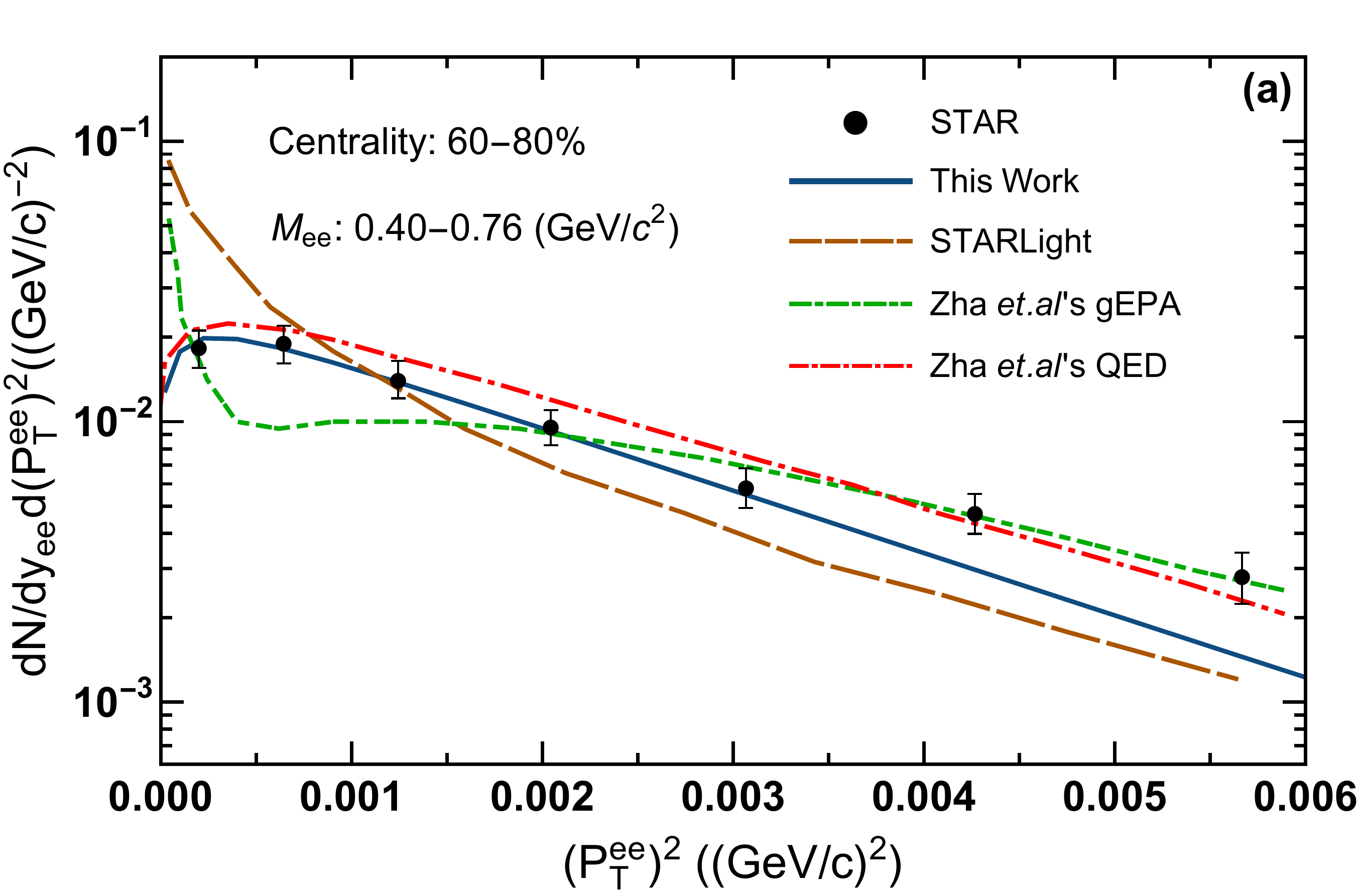}\includegraphics[scale=0.3]{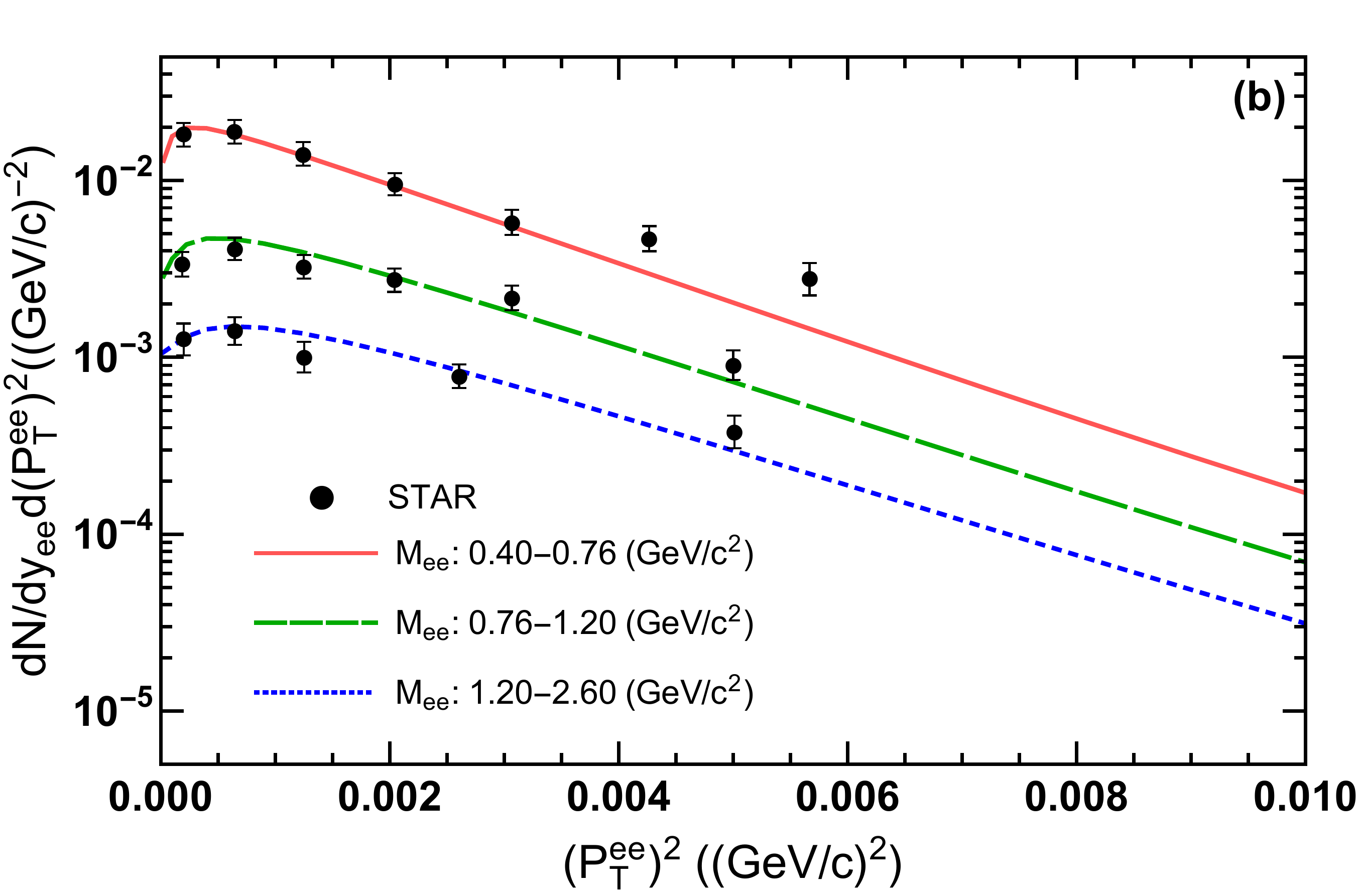}

\caption{The spectra of $(P_{T}^{ee})^{2}$ in 60-80\% centrality. (a) The
STAR data \citep{Adam:2018tdm} and results from some models with
the invariant mass $M_{ee}$ in the range {[}0.40,0.76{]} GeV; (b)
The results of this work with $M_{ee}$ in other ranges. The rapidity
$y_{ee}$ and pseudo-rapidity $\eta_{e}$ are in the range $[-1,1]$
and the transverse momentum of the single electron or positron is
larger than 0.2 GeV. \label{fig:Pee}}
\end{figure}



For the sake of clarity, we define

\begin{align}
\mathbf{P}_{T}^{l\overline{l}} & =\mathbf{k}_{1T}+\mathbf{k}_{2T},\nonumber \\
\mathbf{K}_{T}^{l\overline{l}} & =\frac{1}{2}(\mathbf{k}_{2T}-\mathbf{k}_{1T}),\label{eq:PT_KT_def_01}
\end{align}
where $\mathbf{k}_{1T}$ and $\mathbf{k}_{2T}$ are the transverse
momentum of the lepton and antilepton in the collision (\ref{eq: Subprocess}),
respectively. We can further define $k_{1T}^{l}\equiv|\mathbf{k}_{1T}|$,
$k_{2T}^{\overline{l}}\equiv|\mathbf{k}_{2T}|$, $P_{T}^{l\overline{l}}\equiv|\mathbf{P}_{T}^{l\overline{l}}|$,
$K_{T}^{l\overline{l}}\equiv|\mathbf{K}_{T}^{l\bar{l}}|$, $\varphi$
as the azimuthal angle between the $\mathbf{P}_{T}^{l\overline{l}}$
and $\mathbf{K}_{T}^{l\overline{l}}$, $\varphi_{ll}$ as the azimuthal
angle of $\mathbf{P}_{T}^{l\overline{l}}$, and $\varphi_{l}$ as
the azimuthal angle of $\mathbf{k}_{2T}$. We use $M_{l\overline{l}}$
and $y_{l\overline{l}}$ to denote the invariant mass and rapidity
of the lepton pair respectively, and use $y_{l}$ ($y_{\overline{l}}$)
and $\eta_{l}$ ($\eta_{\overline{l}}$) to denote the rapidity and
pseudo-rapidity of $l$ ($\overline{l}$) respectively.

Following STAR experiments \citep{Adam:2018tdm,Adam:2019mby}, we
set the momentum cutoffs: $k_{1T}^{e^{-}},k_{2T}^{e^{+}}\geq$200
MeV, $P_{T}^{ee}\leq$100 and 150 MeV in UPC and peripheral collisions
respectively, $k_{1}^{\mu^{-}},k_{2}^{\mu^{+}}\in[180,300]$ MeV,
and $P_{T}^{\mu\mu}\leq$100 MeV. The rapidity ranges for $y_{ee}$,
$\eta_{e^{-}}$ and $\eta_{e^{+}}$ are set to $[-1,1]$, and those
for $y_{\mu\mu}$, $\eta_{\mu^{-}}$ and $\eta_{\mu^{+}}$ are set
to $[-0.8,0.8]$.

In Table \ref{tab: Centralities}, we list the centralities and impact
parameters in Au+Au collisions following Ref. \citep{Klein:2018cjh}.


\subsection{Transverse momentum distribution \label{subsec:Transverse-momentum-distribution}}

\begin{figure}[t]
\includegraphics[scale=0.35]{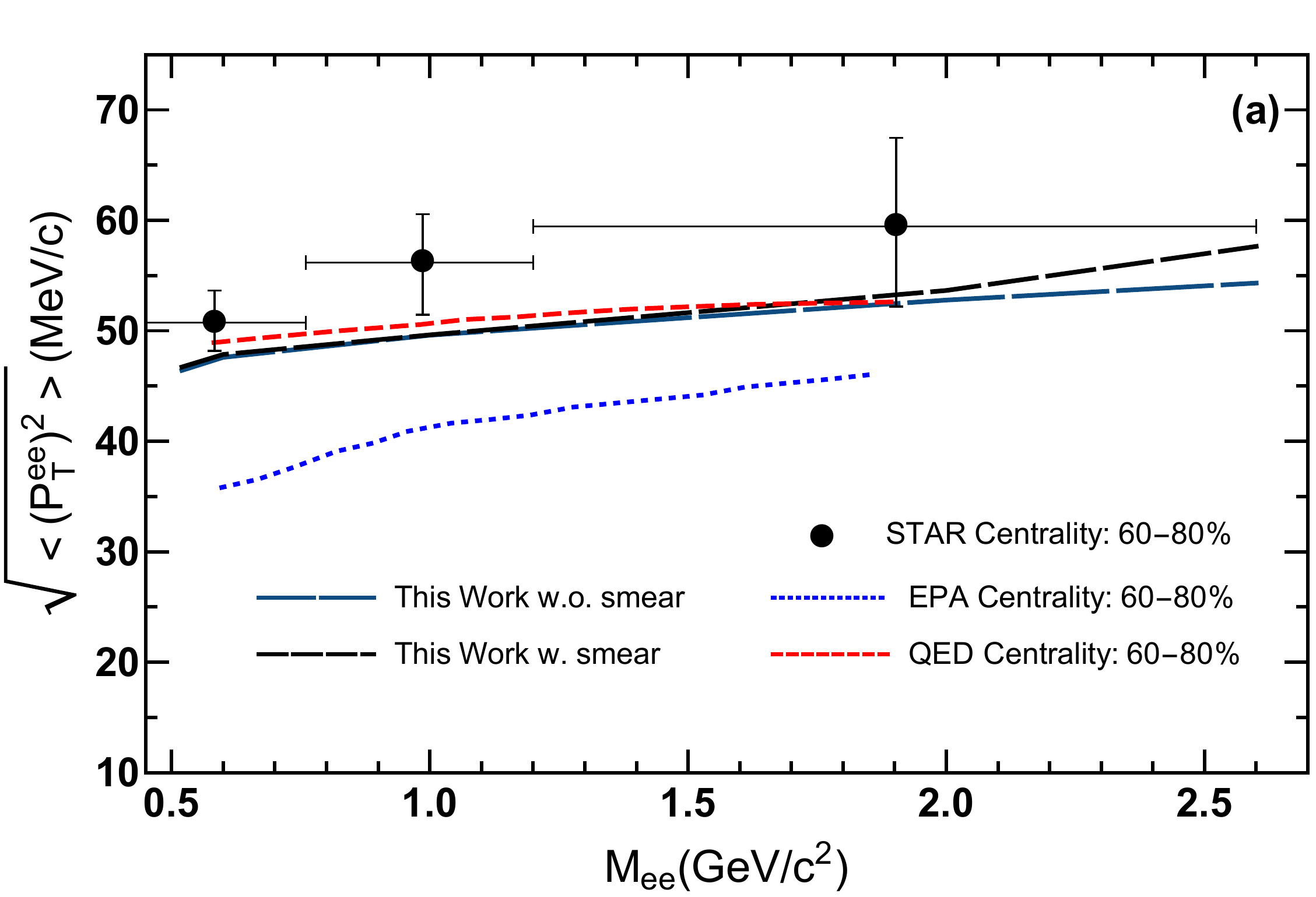} \includegraphics[scale=0.35]{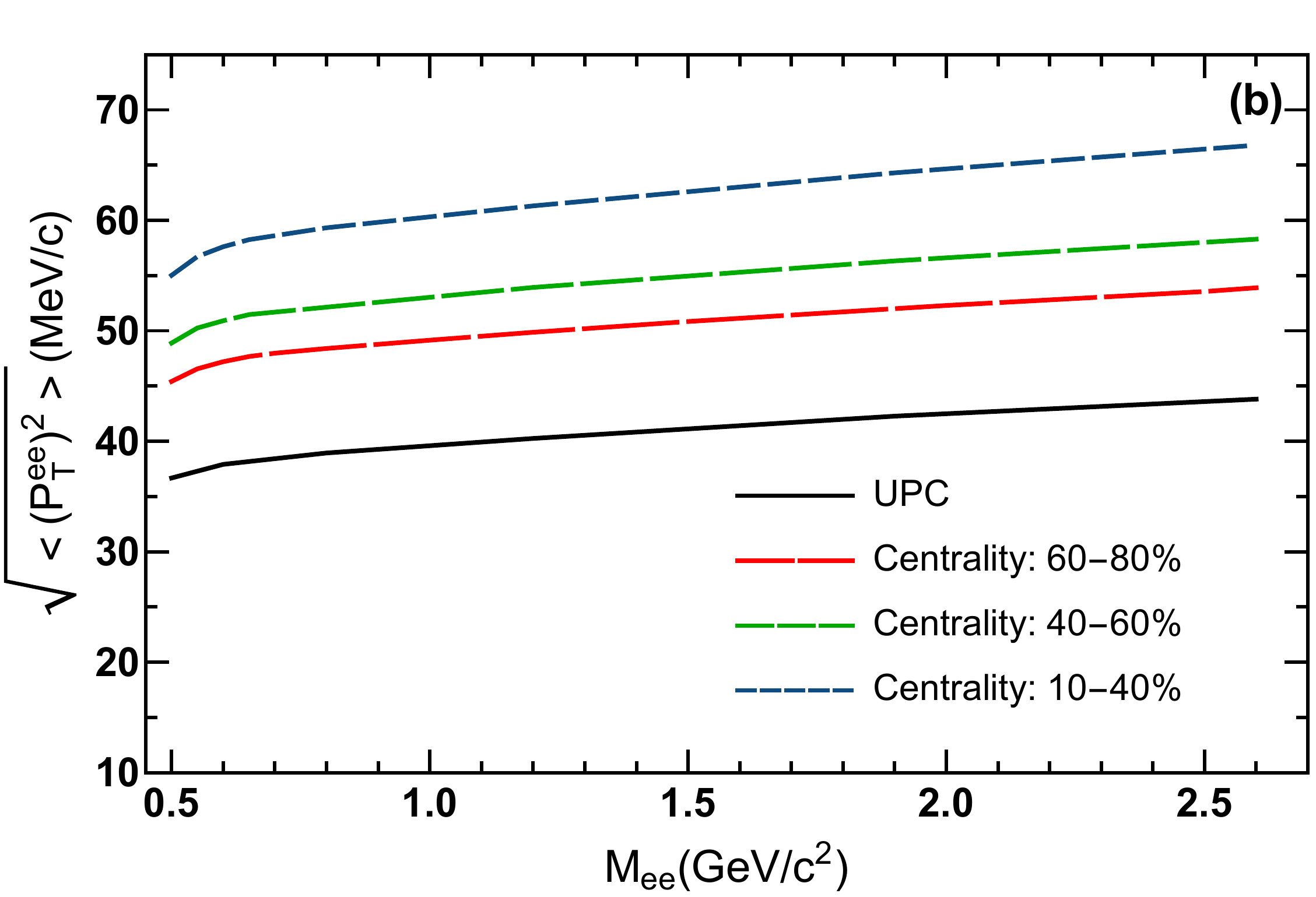}\caption{\label{fig:Pt-Mee-b} The distributions of $\sqrt{\left\langle (P_{T}^{ee})^{2}\right\rangle }$
as functions of $M_{ee}$. (a) The results computed from Eq. (\ref{eq:main-cross-section})
with and without smear corrections. The results from EPA \citep{Adam:2018tdm}
and QED model \citep{Brandenburg:2021lnj} as well as STAR data \citep{Adam:2018tdm}
are also shown; (b) Comparison of results in different centralities.
Other parameters are chosen to be the same as in Fig. \ref{fig:Pee}.}
\end{figure}


\begin{figure}[t]
\includegraphics[scale=0.35]{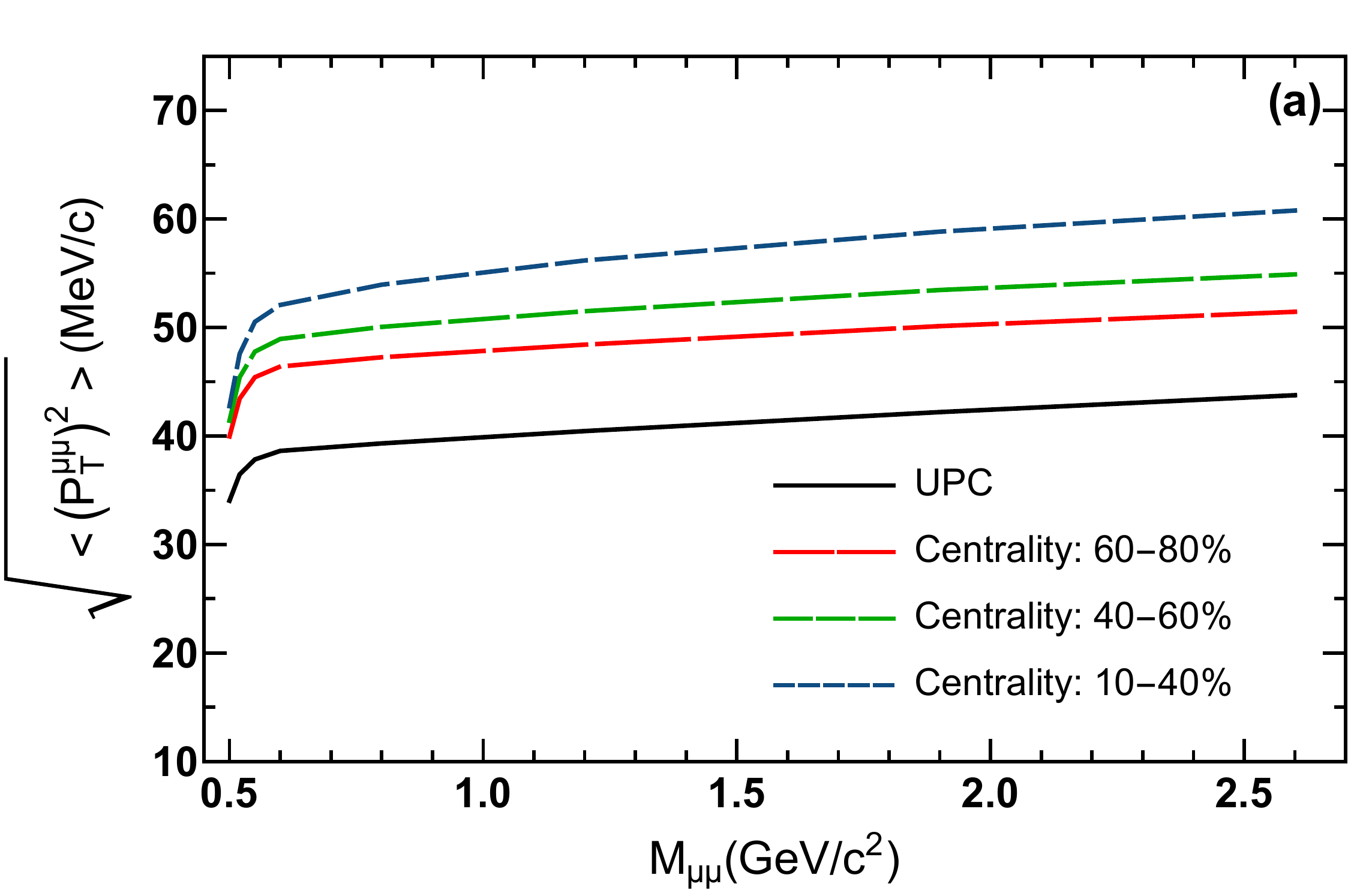} \includegraphics[scale=0.45]{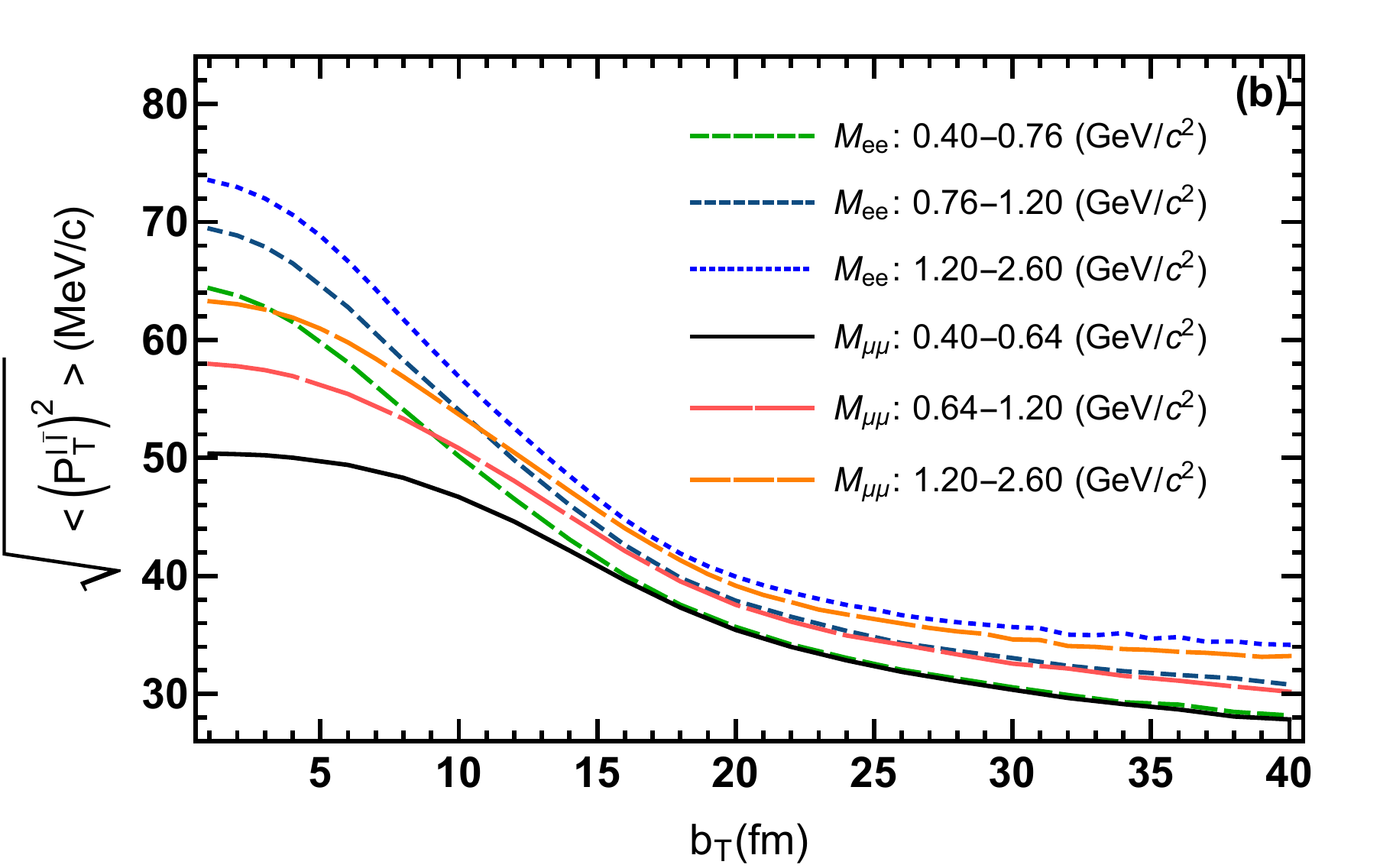}\caption{(a) $\sqrt{\left\langle (P_{T}^{\mu\mu})^{2}\right\rangle }$ as functions
of $M_{\mu\mu}$ in different centralities. (b) $\sqrt{\left\langle (P_{T}^{ee})^{2}\right\rangle }$
and $\sqrt{\left\langle (P_{T}^{\mu\mu})^{2}\right\rangle }$ as functions
of $b_{T}$ in different invariant mass ranges. The parameters for
electrons are chosen to be the same as in Fig. \ref{fig:Pee}. Both
$y_{\mu\mu}$ and $\eta_{\mu}$ are integrated over the range $[-0.8,0.8]$,
and the ranges of $P_{T}^{ee}$ and $P_{T}^{\mu\mu}$ are set to $P_{T}^{ee}\in[0.00,0.15]$
GeV and $P_{T}^{\mu\mu}\in[0.00,0.10]$ GeV. \label{fig:Pt-b}}
\end{figure}


In Fig. \ref{fig:Pee}, we plot the spectra of $(P_{T}^{ee})^{2}$
in 60-80\% centrality in Au+Au collisions at 200 GeV in different
ranges of the invariant mass $M_{ee}$. In the figure we use $dN/dy_{ee}d(P_{T}^{ee})^{2}$,
it can also be expressed as $d\sigma/dy_{ee}d(P_{T}^{ee})^{2}$ through
the relation between the number of final particles $N$ and the cross
section as

\begin{equation}
dN=\frac{d\sigma}{\pi(b_{T,\mathrm{max}}^{2}-b_{T,\mathrm{min}}^{2})}.\label{eq:dN_dsigma}
\end{equation}
In Fig. \ref{fig:Pee}(a), we see that both our result from Eq. (\ref{eq:main-cross-section})
and that based on QED model \citep{Zha:2018tlq} match the experimental
data well especially for small $(P_{T}^{ee})^{2}$. We have numerically
check that the difference between our results and the one based on
QED model \citep{Zha:2018tlq} comes from the choice of the radius
of nuclei $R_{A}$ in the region $(P_{T}^{ee})^{2}\geq0.004$ $\textrm{GeV}^{2}$.
The result of STARLight \citep{Klein:2016yzr} based on EPA and that
from gEPA \citep{Zha:2018tlq} cannot reproduce the data in the small
$(P_{T}^{ee})^{2}$ region because the information about the transverse
momentum and polarization of photons is missing in STARLight and gEPA
which is essential to reproduce the data. We can see in Fig. \ref{fig:Pee}(b)
that the spectra of $(P_{T}^{ee})^{2}$ decrease with increasing $M_{ee}$,
which agrees with the observation in our previous work \citep{Wang:2021kxm}
and experimental data \citep{Adam:2018tdm,Adam:2019mby}.


In Fig. \ref{fig:Pt-Mee-b}, we plot the distribution of $\sqrt{\left\langle (P_{T}^{ee})^{2}\right\rangle }$
as functions of $M_{ee}$ in different centralities. In Fig. \ref{fig:Pt-Mee-b}(a),
we compare the results from Eq. (\ref{eq:main-cross-section}) with
and without smear corrections, those from EPA \citep{Adam:2018tdm}
and QED models \citep{Brandenburg:2021lnj} as well as experimental
data in 60-80\% centrality. The EPA (or STARLight) \citep{Adam:2018tdm}
does not include the information for the transverse momentum of photons
and impact parameters, so it fails to give the broadening effects.
Since the transverse momentum and impact parameter of photons are
incorporated into Eq. (\ref{eq:main-cross-section}), our results
can describe the experimental data \citep{Adam:2018tdm} and are consistent
with recent calculations based on QED models \citep{Brandenburg:2021lnj}.


Another possible missing correction in our previous work \citep{Wang:2021kxm}
and in Eq. (\ref{eq:main-cross-section}) is the smear correction
\citep{STAR:2005gfr}. The limitation of momentum resolution in the
detectors and the Bremsstrahlung of leptons inside the detectors may
change the final transverse momentum distribution of leptons. In experiments,
one needs to add the smear corrections to adjust the $P_{T}^{l\overline{l}}$
distribution in the end. In the current work, we implement the smear
corrections extracted from the measurements of $J/\psi\rightarrow l\overline{l}$
\citep{STAR:2005gfr}. In Fig. \ref{fig:Pt-Mee-b}(a), we find that
the smear corrections lead to an enhancement of $\sqrt{\left\langle (P_{T}^{ee})^{2}\right\rangle }$
when $M_{ee}>2$ GeV and are almost negligible in small $M_{ee}$
region. However, it seems that the smear corrections are still insufficient
to match the data, so higher order corrections beyond the Born level
or medium effects need to be considered in the future. Since the smear
corrections do not have significant effects, we do not include them
into other quantities except $\sqrt{\left\langle (P_{T}^{ee})^{2}\right\rangle }$.


In Fig. \ref{fig:Pt-Mee-b}(b), we observe the broadening of $P_{T}^{ee}$
characterized by $\sqrt{\left\langle (P_{T}^{ee})^{2}\right\rangle }$
grows with decreasing of the impact parameter or the centrality \citep{Adam:2018tdm}.
We note that a similar broadening effect also exists for $P_{T}^{\mu\mu}$
as shown in Fig. \ref{fig:Pt-b}(a). In Fig. \ref{fig:Pt-b}(b), we
plot $\sqrt{\left\langle (P_{T}^{ee})^{2}\right\rangle }$ and $\sqrt{\left\langle (P_{T}^{\mu\mu})^{2}\right\rangle }$
as functions of $b_{T}$ at different values of $M_{ee}$ and $M_{\mu\mu}$.
Again, we observe that $\sqrt{\left\langle (P_{T}^{l\bar{l}})^{2}\right\rangle }$
increases as $b_{T}$ decreases or $M_{l\overline{l}}$ increases.
We see in Fig. \ref{fig:Pt-b} that the results for $e^{+}e^{-}$
and $\mu^{+}\mu^{-}$ are quite similar except that $\sqrt{\left\langle (P_{T}^{\mu\mu})^{2}\right\rangle }$
is less than $\sqrt{\left\langle (P_{T}^{ee})^{2}\right\rangle }$
at fixed $b_{T}$ and $M_{l\overline{l}}$ due to the mass effect.


\subsection{Azimuthal angle distributions \label{subsec:Azimuthal-angle-distribution}}

\begin{figure}[t]
\includegraphics[scale=0.35]{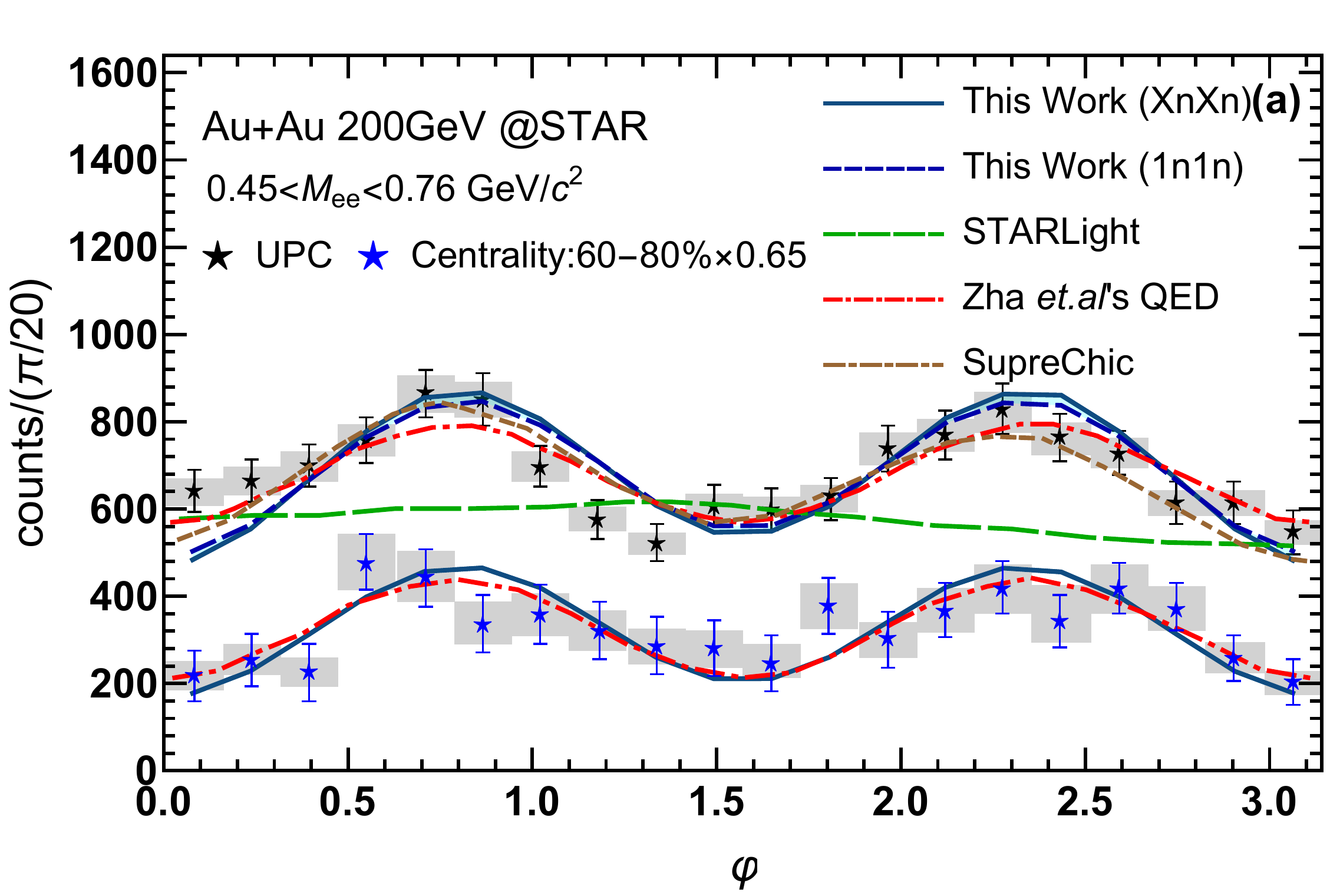}\includegraphics[scale=0.35]{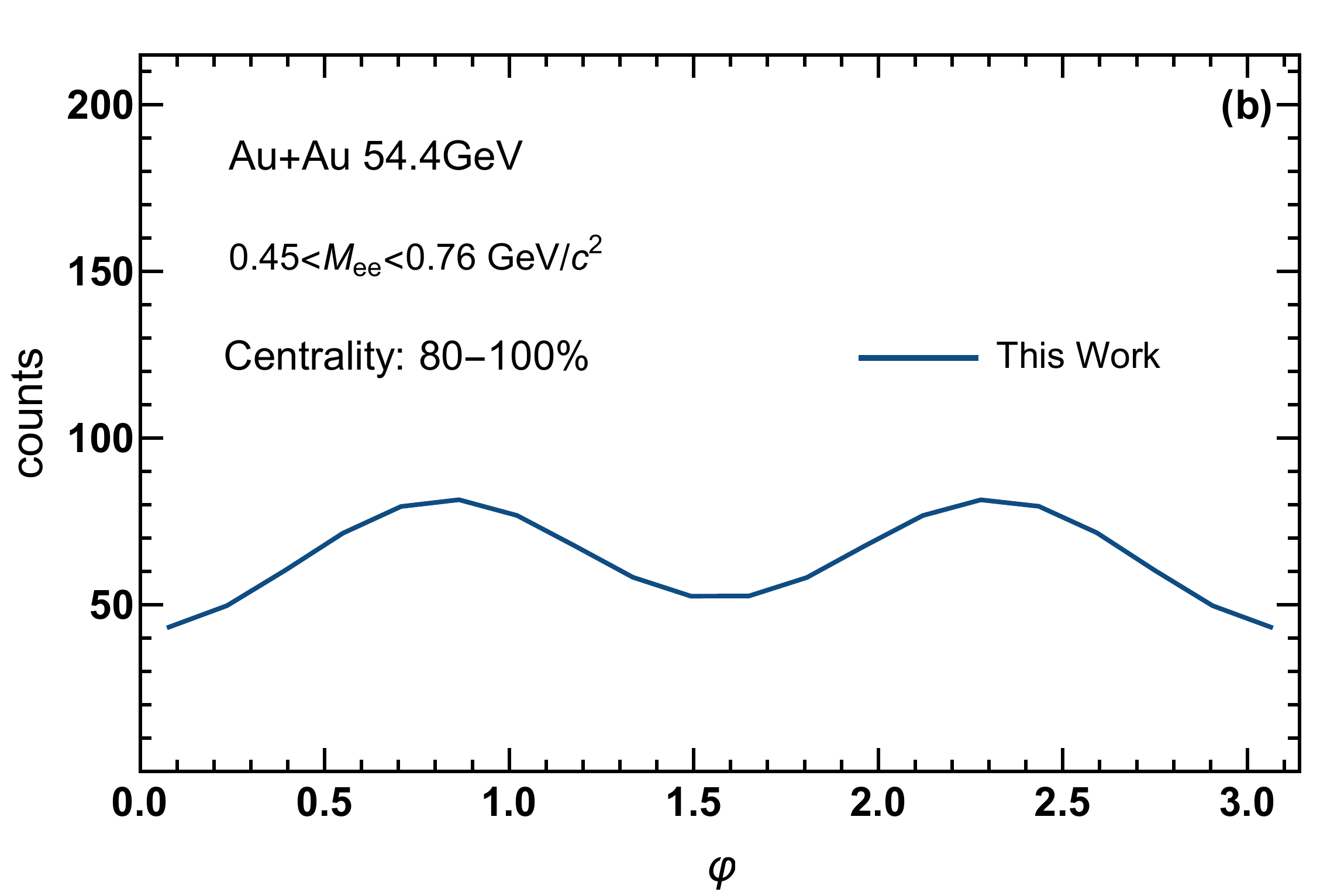}\caption{The distributions of $\varphi$ in Au+Au collisions. The invariant
mass of electron pairs is in the range $M_{ee}\in${[}0.45,0.76{]}
GeV. (a) UPC and 60-80\% centrality at 200 GeV. The dark-blue-solid,
dark-blue-dashed, green-dashed, red-dash-dotted and brown-dash-dotted
lines denote the results from Eq. (\ref{eq:main-cross-section}) for
emission of $X>1$ neutrons, those for emission of one neutron by
each colliding nucleus, STARLight \citep{Adam:2019mby}, QED model
\citep{Zha:2018tlq} and SupreChic program \citep{Adam:2019mby},
respectively. The data points are from STAR measurements \citep{Adam:2019mby}.
(b) Prediction for 80-100\% centrality in Au+Au collisions at 54.4
GeV. 
\label{fig:Angle-e}}
\end{figure}


\begin{figure}
\includegraphics[scale=0.35]{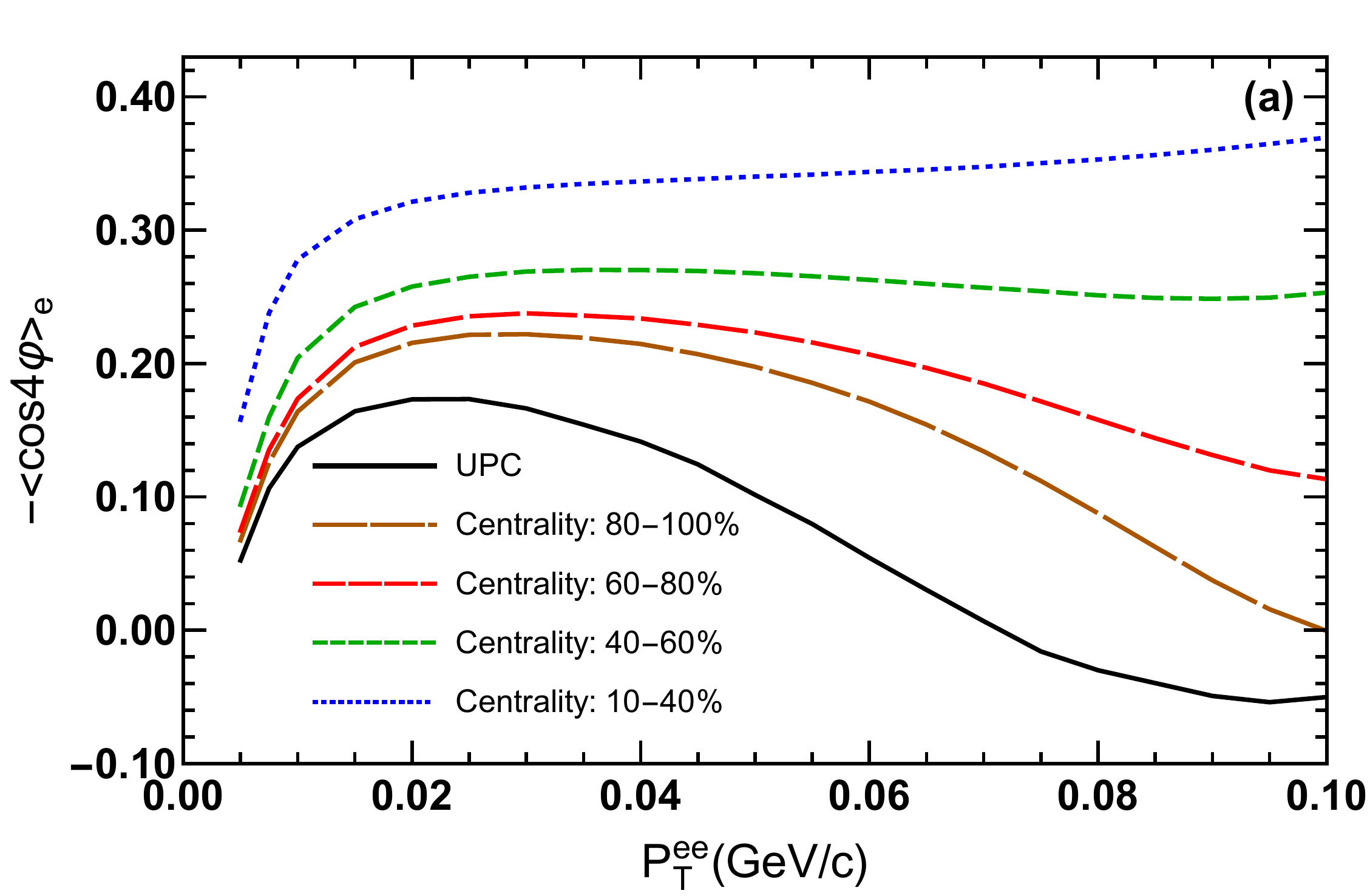} \includegraphics[scale=0.35]{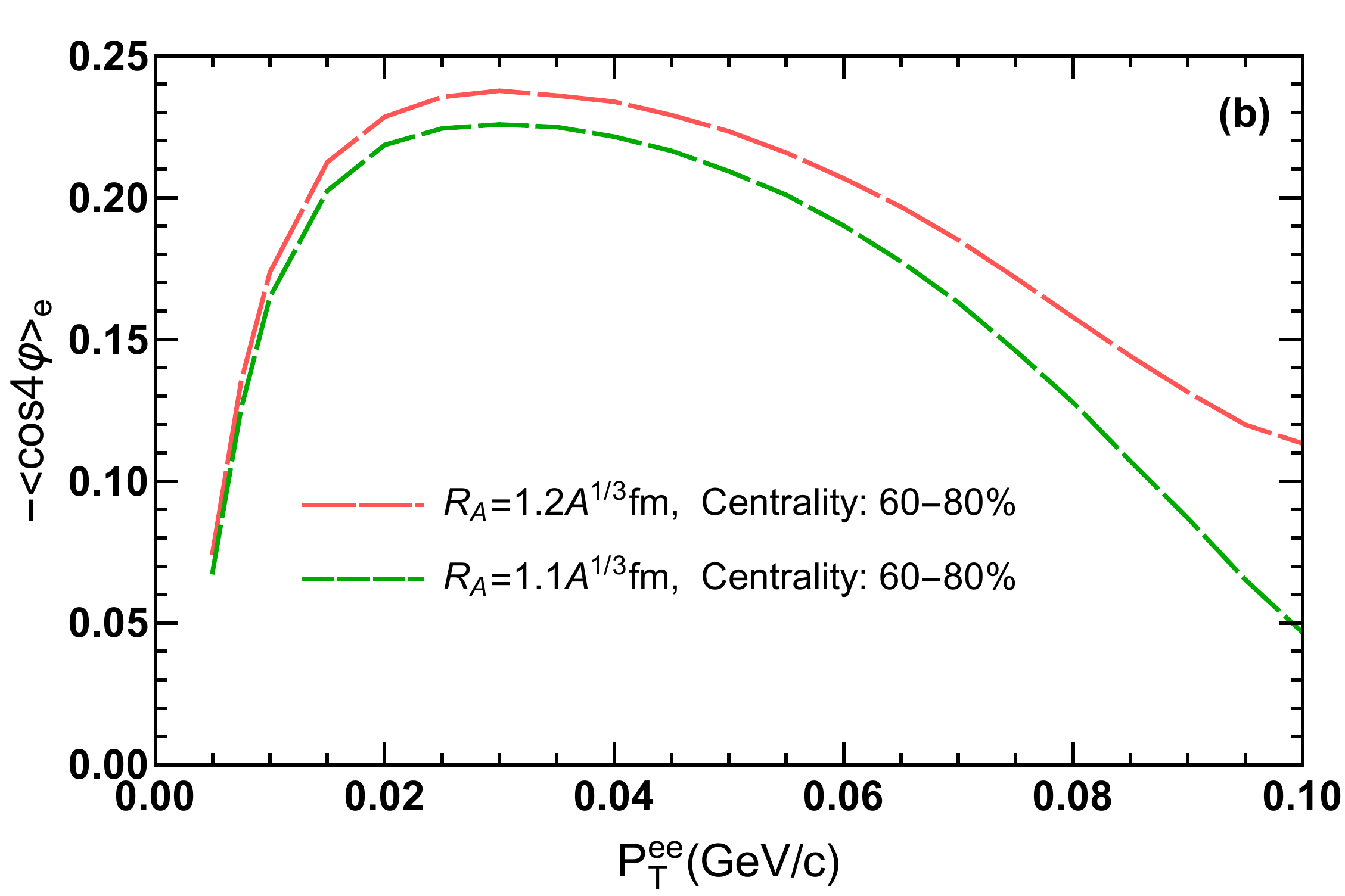}\caption{The average of $\cos(4\varphi)$ for electron pairs as functions of
$P_{T}^{ee}$ in Au+Au collisions at 200 GeV. (a) In different centrality
bins; (b) In 60-80\% centrality with different nucleus radius $R_{A}$.
The ranges for $y_{e^{-}}$ and $y_{e^{+}}$ are $[-1,1]$, and the
range for $K_{T}^{ee}$ is {[}0.20, 0.40{]} GeV. \label{fig:4Angle-P-e}}
\end{figure}


\begin{figure}
\centering{}\includegraphics[scale=0.3]{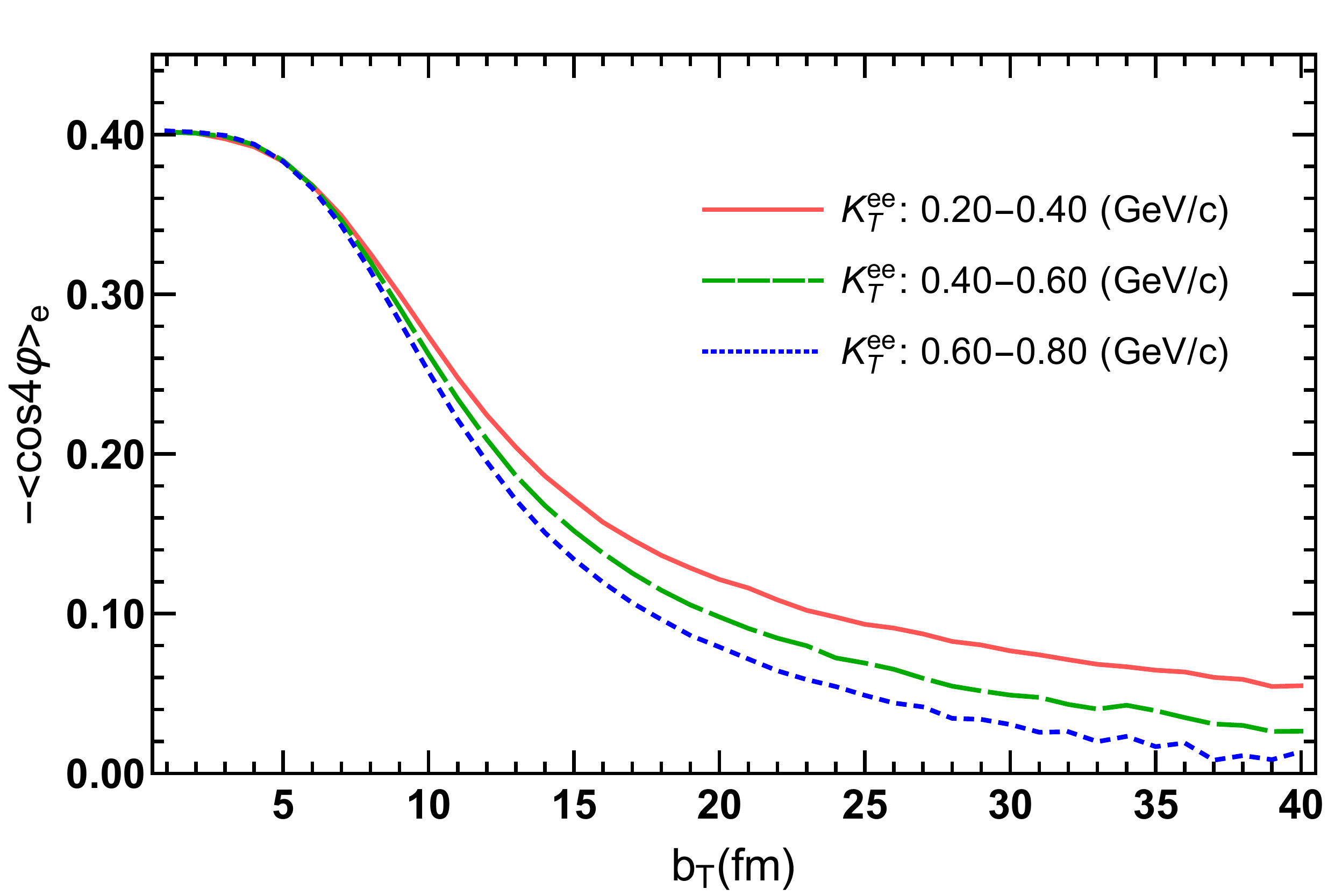}\caption{The average of $\cos(4\varphi)$ for electron pairs as functions of
$b_{T}$ in Au+Au collisions at 200 GeV within different $K_{T}^{ee}$
bins. The ranges for $y_{ee}$ and $P_{T}^{ee}$ are $[-1,1]$ and
{[}0.00,0.15{]} GeV, respectively. \label{fig:4Angle-b-e}}
\end{figure}


\begin{figure}
\centering{}\includegraphics[scale=0.35]{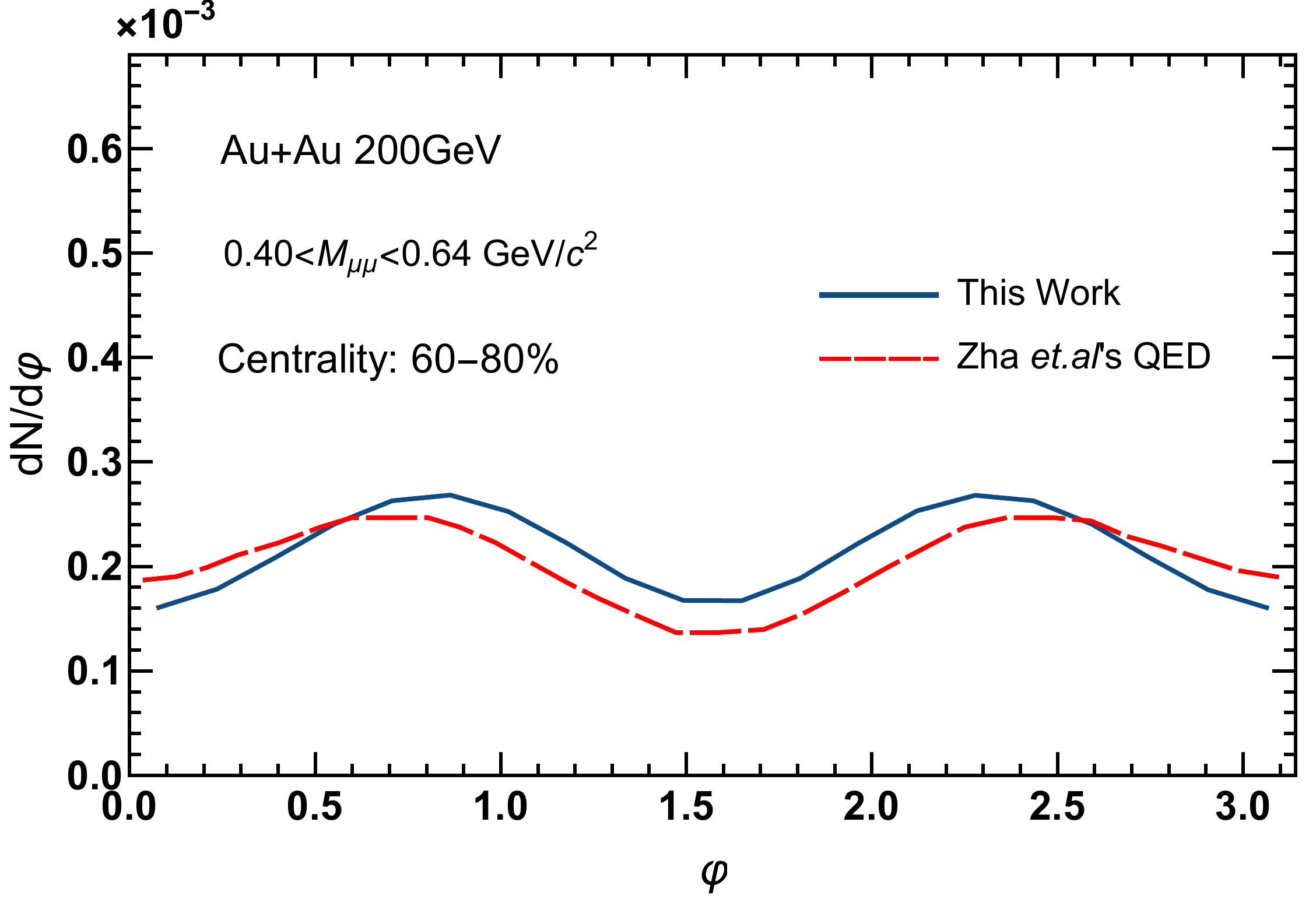}\caption{The distributions of $\varphi$ in Au+Au collisions at 200 GeV and
60-80\% centrality. The range of $M_{\mu\mu}$ is set to {[}0.40,0.64{]}
GeV. The dark-blue-solid and red-dashed lines are the results from
Eq.(\ref{eq:main-cross-section}) and QED models \citep{Zha:2018tlq,Zhou:2022gbh},
respectively. \label{fig:Angle-u}}
\end{figure}


\begin{figure}
\includegraphics[scale=0.35]{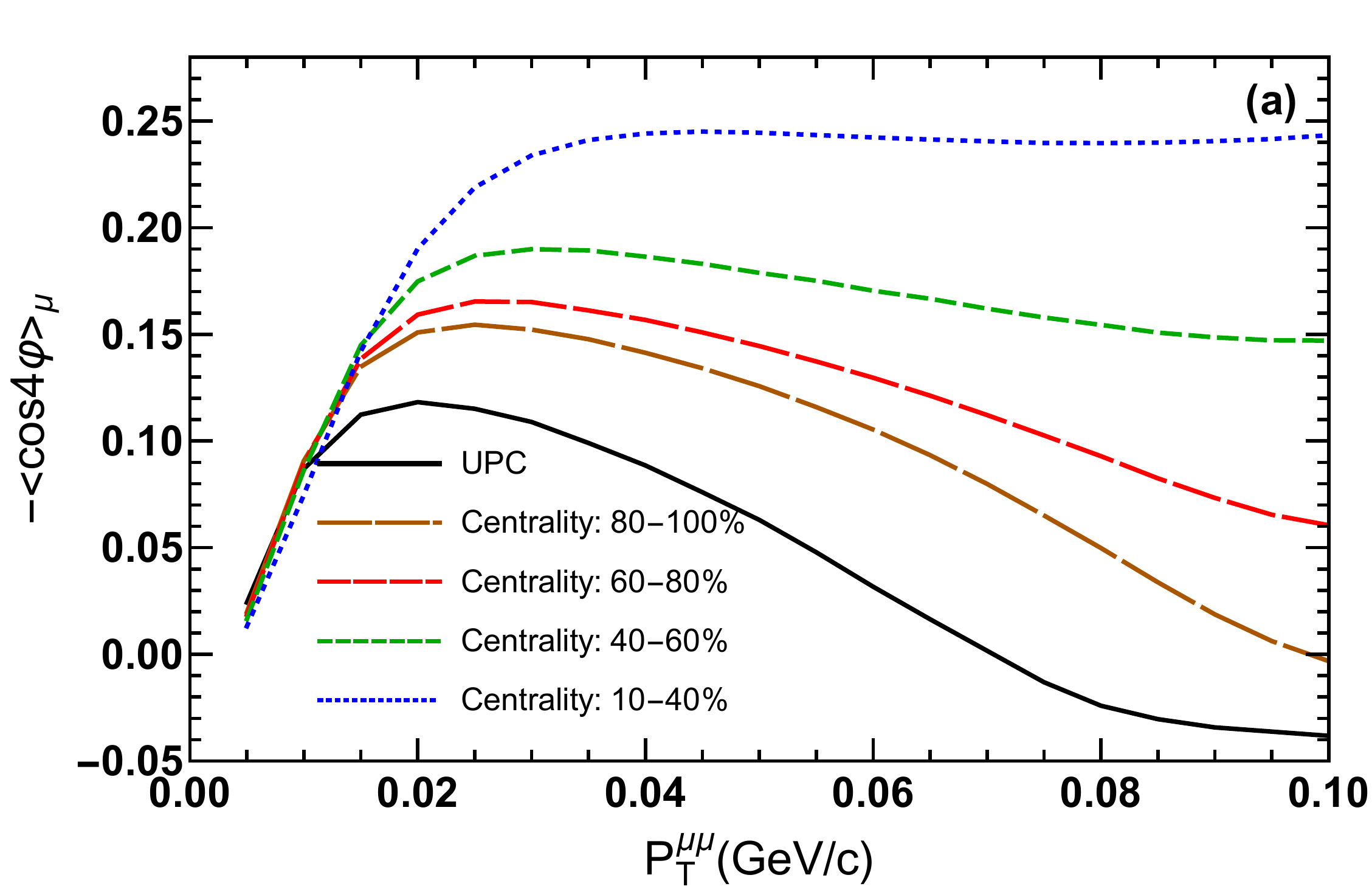} \includegraphics[scale=0.35]{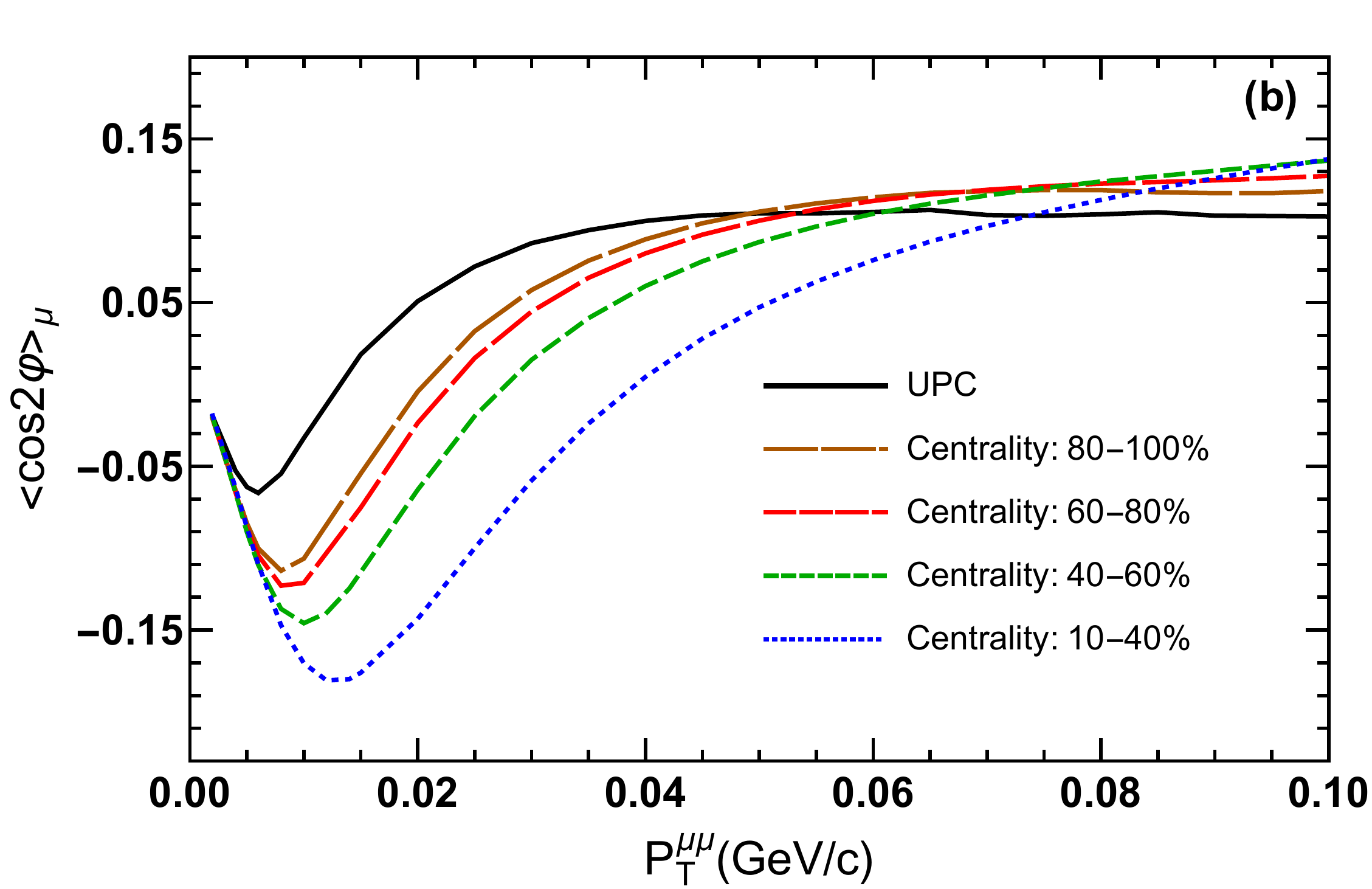}\caption{The distributions of (a) $-\cos(4\varphi)$ and (b) $\cos(2\varphi)$
for muon pairs as functions of $P_{T}^{\mu\mu}$ in Au+Au collisions
at 200 GeV in different centrality bins. The ranges for $y_{\mu^{-}}$
and $y_{\mu^{+}}$ are $[-0.8,0.8]$, and the range for $K_{T}^{\mu\mu}$
is {[}0.2, 0.4{]} GeV. \label{fig:4 and 2Angle-P-mu}}
\end{figure}


\begin{figure}
\includegraphics[scale=0.32]{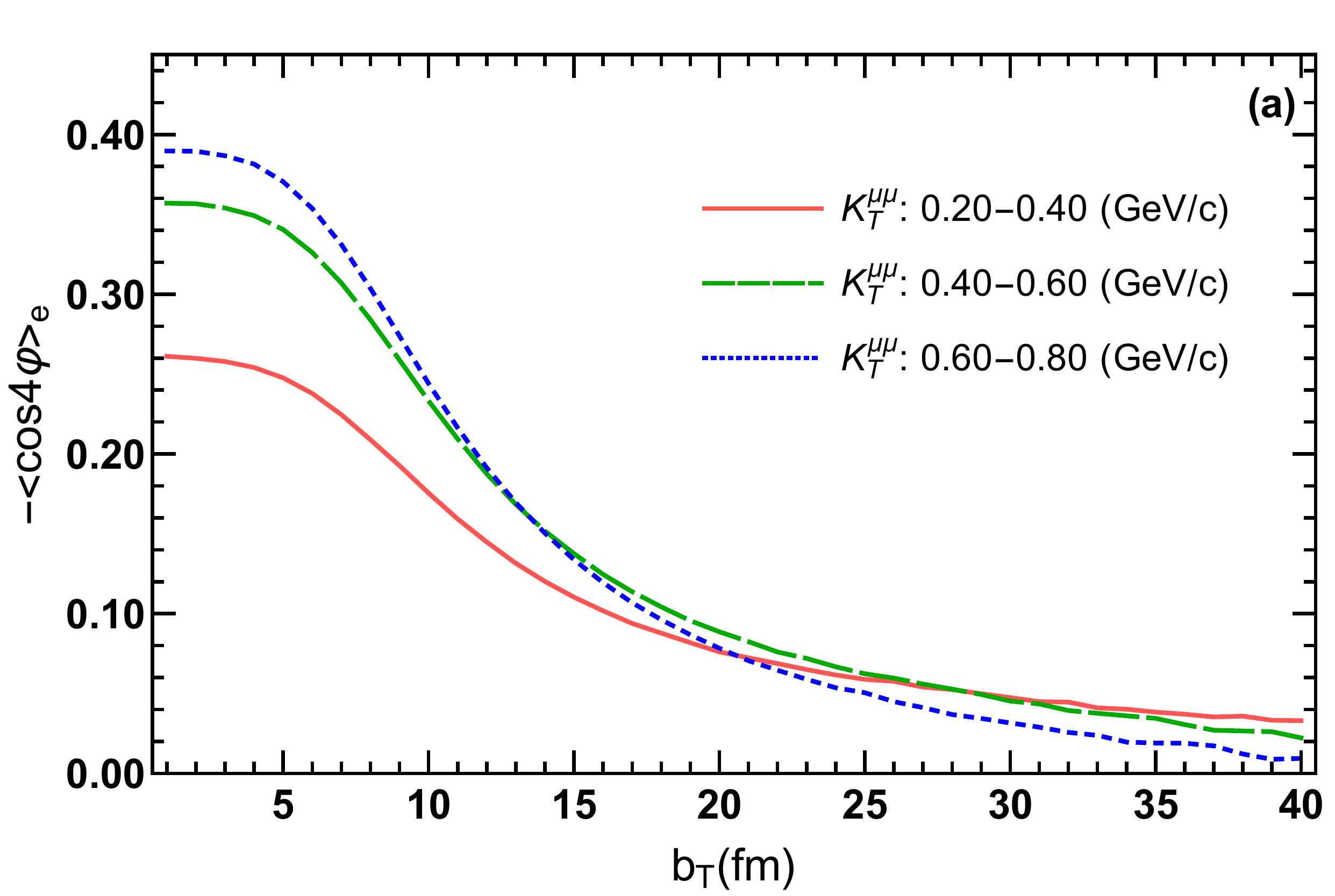}\includegraphics[scale=0.33]{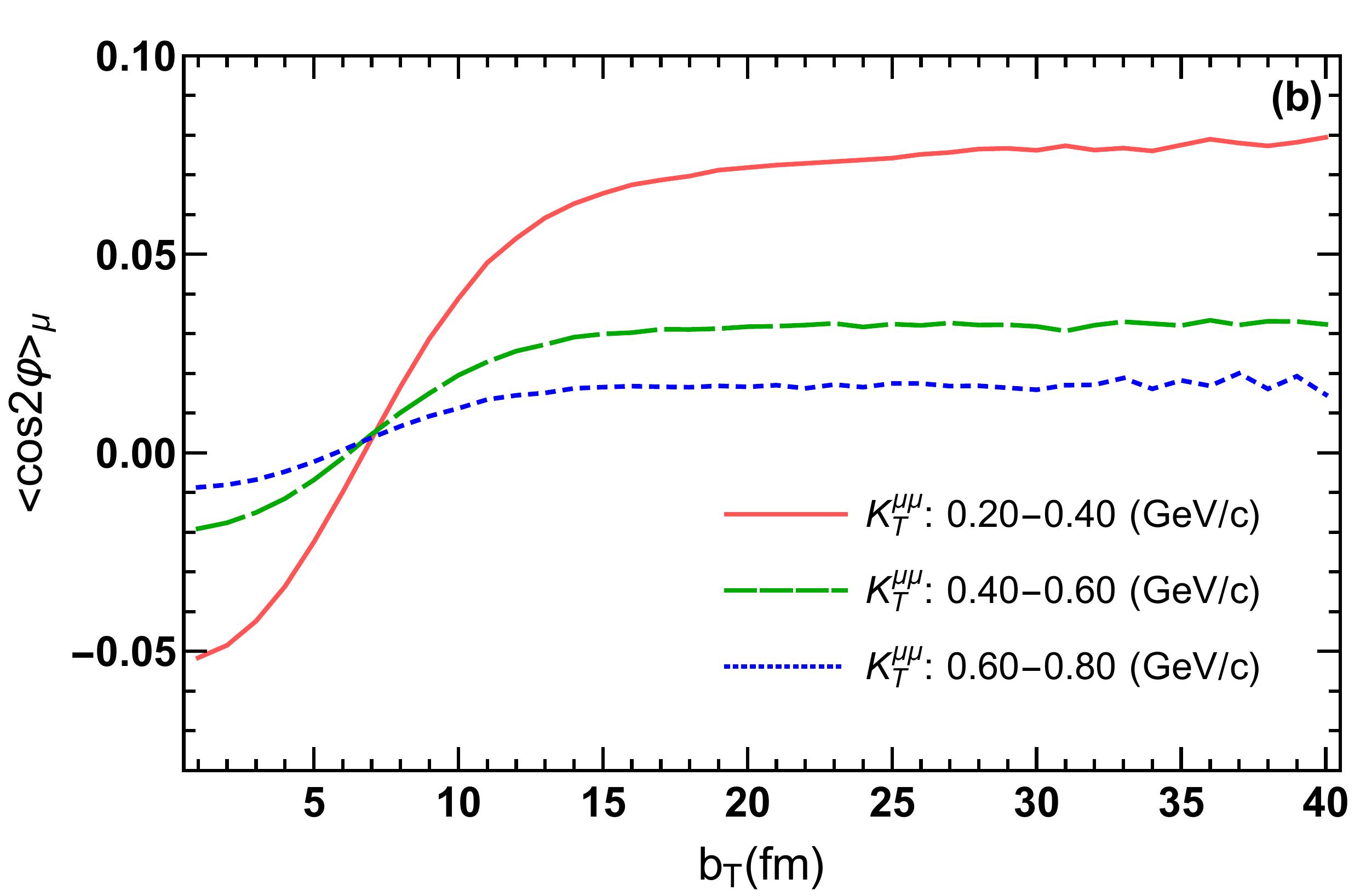}\caption{The distributions of (a) $-\cos(4\varphi)$ and (b) $\cos(2\varphi)$
for muon pairs as functions of $b_{T}$ in Au+Au collisions at 200
GeV for different $K_{T}^{\mu\mu}$. The ranges for $y_{\mu^{-}}$
and $y_{\mu^{+}}$ are $[-0.8,0.8]$, and the range for $P_{T}^{\mu\mu}$
is {[}0.00, 0.10{]} GeV. \label{fig:4 and 2Angle-b-mu}}
\end{figure}


In Fig. \ref{fig:Angle-e}, we plot the azimuthal angle $\varphi$
distributions for electron pairs in Au+Au collisions. We show the
results for UPC and 60-80\% centrality at 200 GeV and those for 80-100\%
centrality at 54.4 GeV in comparison with STAR data \citep{Adam:2019mby}.


In Fig. \ref{fig:Angle-e}(a), we find that our results agree with
experimental data and also close to the results of the QED model \citep{Zha:2018tlq}.
Again, due to the lack of the essential information for the transverse
momentum and polarization of photons, the STARLight results \citep{Adam:2019mby}
do not match the data. It is another piece of evidence that our key
formula (\ref{eq:main-cross-section}) which includes the spectra
of transverse momentum and polarization of photons can better describe
the experiments than EPA.


We see in Fig. \ref{fig:Angle-e} that there are a $\cos(4\varphi)$
modulation behavior \citep{Li:2019sin,Li:2019yzy}. Such a modulation
behavior is a signal for the linear polarization of incoming photons
and is connected with the vacuum birefringence \citep{Adam:2019mby}.
The same phenomenon also exists in gluons \citep{Metz:2011wb,Akcakaya:2012si,Pisano:2013cya}.
In Fig. \ref{fig:Angle-e}(b), we present the results for the modulation
behavior in Au+Au collisions at 54.4 GeV and 80-100\% centrality.
Our results from Eq. (\ref{eq:main-cross-section}) show that that
the modulation behavior at 54.4GeV also satisfies the function $1+A_{4\varphi}\cos(4\varphi)$
($A_{4\varphi}$ is the coefficient) similar to the one at 200GeV. 


Now we take a look at the average of $\cos(4\varphi)$ on independent
kinematic variables in Fig. \ref{fig:4Angle-P-e}. We follow Refs.
\citep{Li:2019sin,Li:2019yzy} to define the average of $\cos(4\varphi)$
and $\cos(2\varphi)$ as 
\begin{equation}
\left\langle \cos(n\varphi)\right\rangle =\frac{\int\frac{d\sigma}{d\mathcal{P}.\mathcal{S}.}\cos(n\varphi)d\mathcal{P}.\mathcal{S}.}{\int\frac{d\sigma}{d\mathcal{P}.\mathcal{S}.}d\mathcal{P}.\mathcal{S}.},\;n=2,4.
\end{equation}
where $\mathcal{P}.\mathcal{S}.$ denotes the independent kinematic
variables in collisions. For example, if we want to compute $\left\langle \cos(n\varphi)\right\rangle $
as a function of $P_{T}^{l\overline{l}}$, then $\mathcal{P}.\mathcal{S}.$
stands for all independent kinematic variables except $P_{T}^{l\overline{l}}$,
such as the impact parameter, invariant mass, rapidity, and so on.
In Fig. \ref{fig:4Angle-P-e}(a), we observe that the magnitude of
$\left\langle \cos(4\varphi)\right\rangle _{e}$ decreases with the
centrality. In UPC, we find that $\left\langle \cos(4\varphi)\right\rangle _{e}$
oscillates slowly as a function of $P_{T}^{ee}$. In collisions of
80-100\% and 60-80\% centralities, it reaches the maximum value at
about $P_{T}^{ee}\sim$0.02-0.03 GeV and then decreases. When the
centrality is less than 40\%, $-\left\langle \cos(4\varphi)\right\rangle _{e}$
always increases slowly with $P_{T}^{ee}$. These shapes are also
observed in early works \citep{Li:2019sin,Li:2019yzy}. In Fig. \ref{fig:4Angle-P-e}(b),
we present the $R_{A}$ dependence of $\left\langle \cos(4\varphi)\right\rangle _{e}$.
The nuclear radius $R_{A}$ we use is different from $R_{A}=1.1A^{1/3}$
used in Refs. \citep{Li:2019sin,Li:2019yzy}. We find that $\left\langle \cos(4\varphi)\right\rangle _{e}$
is sensitive to $R_{A}$ when $P_{T}^{ee}\geq0.02$ GeV/$c$. Therefore,
it may be possible to measure $R_{A}$ through $\left\langle \cos(4\varphi)\right\rangle _{e}$
in experiments.


As a comparison with the results of Fig. \ref{fig:4Angle-P-e}, we
plot the impact parameter dependence of $\left\langle \cos(4\varphi)\right\rangle _{e}$
in Fig. \ref{fig:4Angle-b-e}. We see that the magnitude of $\left\langle \cos(4\varphi)\right\rangle _{e}$
decreases with increasing $b_{T}$, consistent with Fig. \ref{fig:4Angle-P-e}(a).
We also see that it decreases slightly with increasing $K_{T}^{ee}$.


We also study the azimuthal angle $\varphi$ distributions for muon
pairs. In Fig. \ref{fig:Angle-u}, we can see that our results from
Eq. (\ref{eq:main-cross-section}) agree with those based on QED models
\citep{Zha:2018tlq}. But our results and those of QED models for
muon pairs are similar to those for electron pairs but the magnitudes
are smaller than the STAR data \citep{Zhou:2022gbh}.


In Fig. \ref{fig:4 and 2Angle-P-mu}(a), we plot $-\left\langle \cos(4\varphi)\right\rangle _{\mu}$
as functions of $P_{T}^{\mu\mu}$ at different centralities. We find
that the shape of $-\left\langle \cos(4\varphi)\right\rangle _{\mu}$
is similar to $-\left\langle \cos(4\varphi)\right\rangle _{e}$ in
Fig. \ref{fig:4Angle-P-e}(a), except in the region of small $P_{T}^{l\overline{l}}$.
When $P_{T}^{l\overline{l}}<0.015$ GeV, $-\left\langle \cos(4\varphi)\right\rangle _{\mu}$
increases with the centrality while $-\left\langle \cos(4\varphi)\right\rangle _{e}$
decreases with the centrality.


The notable difference between $\varphi$ distributions of electron
pairs and muon pairs is in the distribution of $\cos(2\varphi)$.
As pointed out in Refs. \citep{Li:2019sin,Li:2019yzy}, $\left\langle \cos(2\varphi)\right\rangle _{l}$
is proportional to $\left(m_{l}/P_{T}^{l\overline{l}}\right)^{2}$
with $m_{l}$ being the lepton mass. For electrons, we have $\left\langle \cos(2\varphi)\right\rangle _{e}\propto\left(m_{e}/P_{T}^{ee}\right)^{2}\rightarrow0$
since $m_{e}/P_{T}^{ee}\ll1$, while for muons, which are much heavier
than electrons, we have $\left\langle \cos(2\varphi)\right\rangle _{\mu}\propto\left(m_{\mu}/P_{T}^{\mu\mu}\right)^{2}$
which is not negligible. We have numerically checked that $\left\langle \cos(2\varphi)\right\rangle _{e}$
is very close to zero, but $\left\langle \cos(2\varphi)\right\rangle _{\mu}$
is sizable as shown in Fig. \ref{fig:4 and 2Angle-P-mu}(b). We observe
that $\left\langle \cos(2\varphi)\right\rangle _{\mu}$ as functions
of $P_{T}^{\mu\mu}$ has minimal values in the range $P_{T}^{\mu\mu}\sim$0.005-0.015
GeV at different centralities. For large $P_{T}^{\mu\mu}$, $\left\langle \cos(2\varphi)\right\rangle _{\mu}$
beocmes saturated at centralities larger than $60\%$ and in UPC.


In Fig. \ref{fig:4 and 2Angle-b-mu}, we present the distributions
of $-\left\langle \cos(4\varphi)\right\rangle _{\mu}$ and $\left\langle \cos(2\varphi)\right\rangle _{\mu}$
as functions of $b_{T}$ for different $K_{T}^{\mu\mu}$. We see $-\left\langle \cos(4\varphi)\right\rangle _{\mu}$
decreases and $\left\langle \cos(2\varphi)\right\rangle _{\mu}$ increases
with increasing $b_{T}$, which is consistent with the centrality
dependence in Fig. \ref{fig:4 and 2Angle-P-mu}. Compared with the
results for electron pairs in Fig. \ref{fig:4Angle-b-e}, we find
that the magnitude of $\left\langle \cos(4\varphi)\right\rangle _{\mu}$
is smaller than that of $\left\langle \cos(4\varphi)\right\rangle _{e}$
due to the mass effect.


\subsection{Invariant mass distribution \label{subsec:Invariant-mass-distribution}}

\begin{figure}[t]
\includegraphics[scale=0.35]{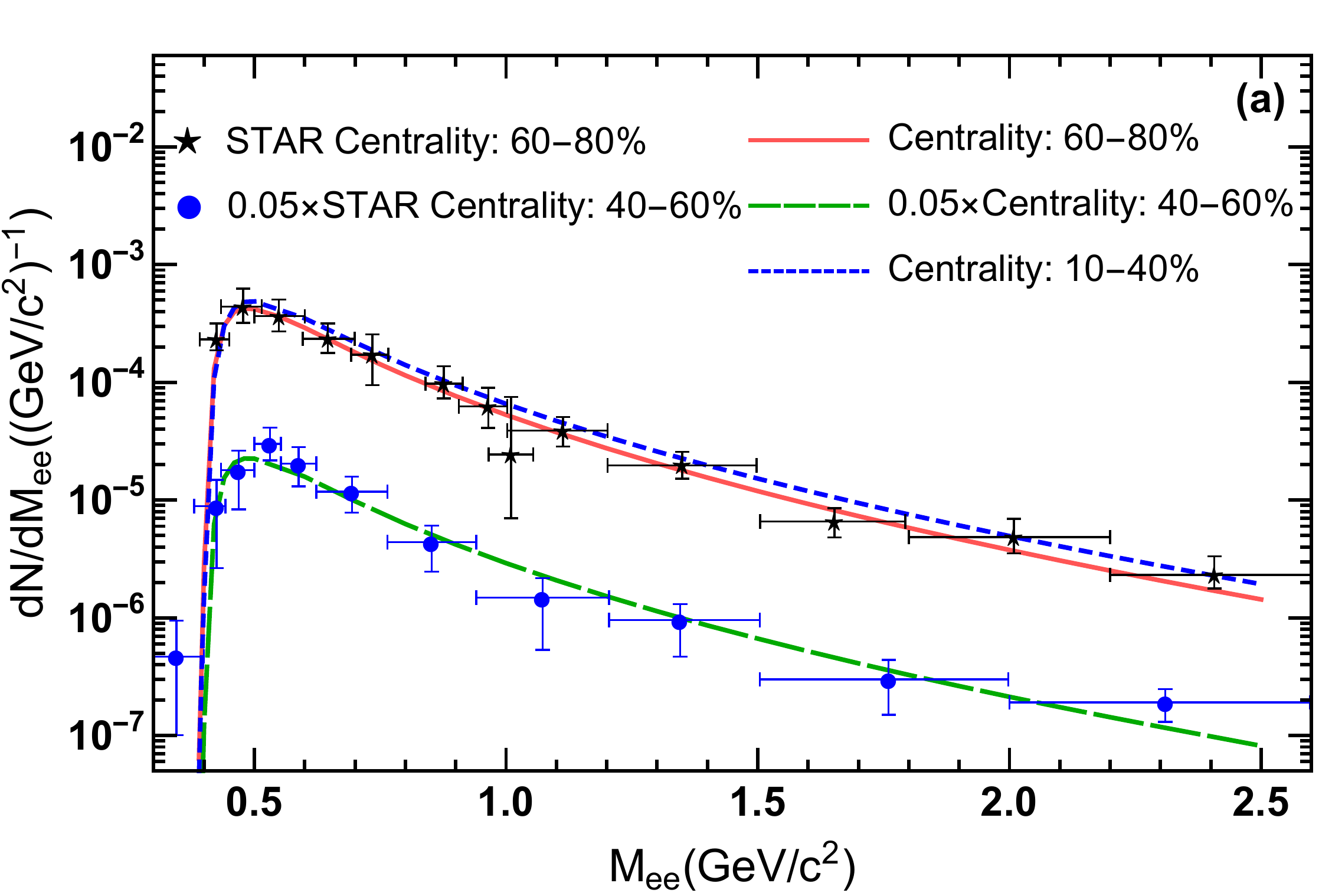}\includegraphics[scale=0.35]{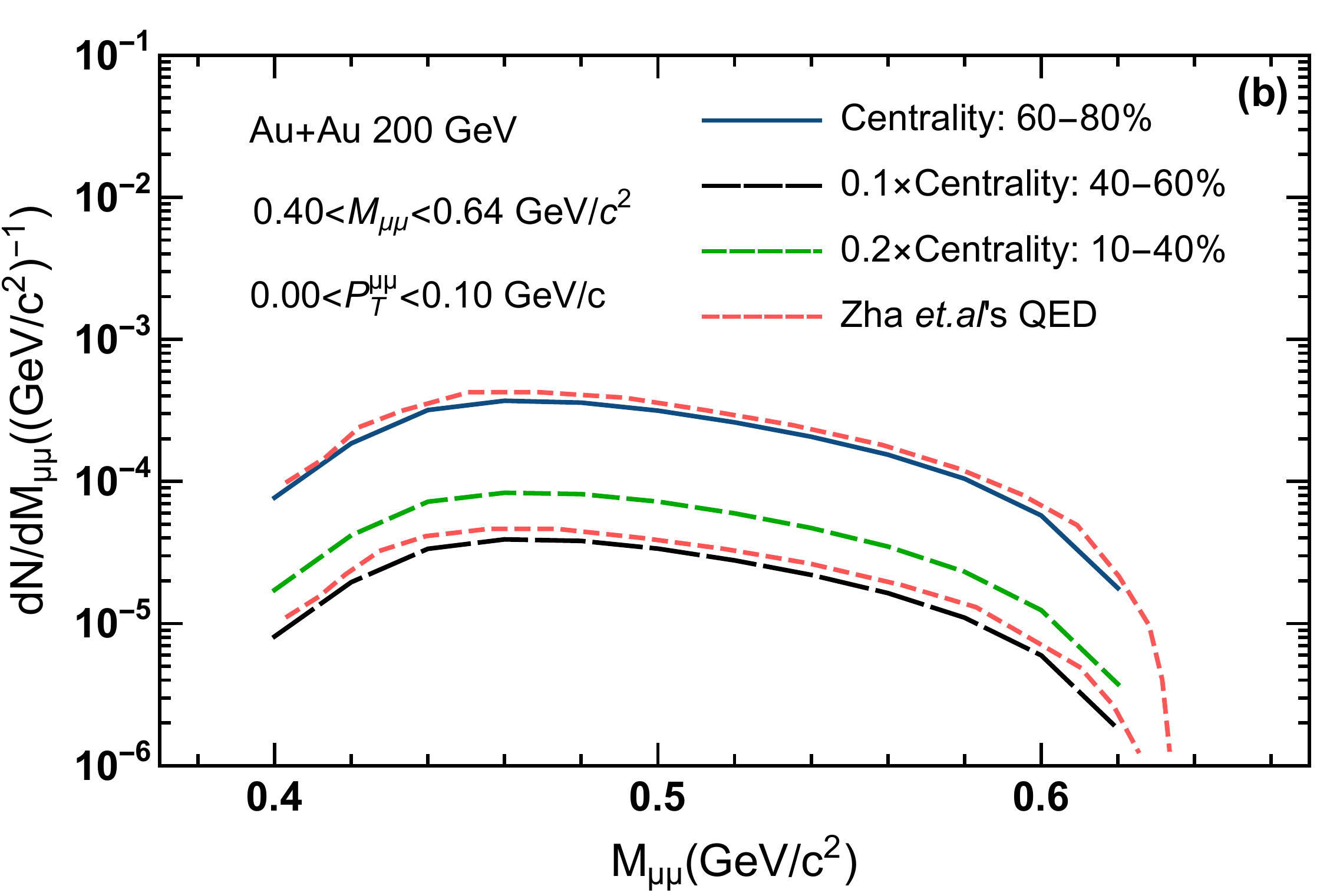}\caption{The invariant mass distributions for (a) electron pairs and (b) muon
pairs in Au+Au collisions at 200 GeV and different centralities. The
data are from STAR measurements \citep{Adam:2018tdm,Zhou:2022gbh}.
The ranges for $y_{ee}$, $\eta_{e^{-}}$ and $\eta_{e^{+}}$ are
$[-1,1]$, and the transverse momentum of the single electron or single
positron is larger than 0.2 GeV. The ranges of $y_{\mu\mu}$, $\eta_{\mu^{-}}$
and $\eta_{\mu^{+}}$ are $[-0.8,0.8]$, and the momentum ranges for
the single muon or single antimuon are 0.18-0.30 GeV. \label{fig:M}}
\end{figure}


In this subsection, we present the invariant mass distributions for
electron and muon pairs in Au+Au collisions at 200 GeV.

We plot in Fig. \ref{fig:M}(a) the results from Eq. (\ref{eq:main-cross-section})
for electron pairs which are in good agreement with the experimental
data \citep{Adam:2018tdm}. The invariant mass distributions for electron
pairs decrease with increasing $M_{ee}$ when $M_{ee}\geq0.5$ GeV
and are not very sensitive to centralities (silghtly decrease with
the centrality). In Fig. \textcolor{blue}{\ref{fig:M}}(b), we present
our results for muon pairs which are also in good agreement with the
experimental data \citep{Zhou:2022gbh}. We find a similar centrality
dependence in the invariant mass distribution for muon pairs. The
difference between our results and the experimental data may come
from higher order corrections, e.g. a possible contribution from Sudakov
factors.


\section{Conclusion and outlook \label{sec:Summary}}

We study the lepton pair photoproduction in peripheral heavy-ion collisions
based on our previous work in Ref. \citep{Wang:2021kxm}. We calculate
the distributions of the transverse momentum $P_{T}^{l\overline{l}}$,
the azimuthal angle $\varphi$ and invariant mass $M_{l\overline{l}}$
for lepton pairs as functions of the impact parameter $b_{T}$ (or
centrality equivalently) and other kinematic variables in Au+Au collisions.


Our results from Eq. (\ref{eq:main-cross-section}) are in good agreement
with the experimental data for electron pairs. The information on
the transverse momentum and polarization for photons is essential
to describe the experimental data. Our results for the azimuthal angle
distribution of muon pairs do not match the experimental data very
well but agree with other models based on QED. Such differences between
theoretical results and experimental data require further studies
in the muon pair production.


Although our results for $\sqrt{\left\langle (P_{T}^{ee})^{2}\right\rangle }$
in Fig. \ref{fig:Pt-Mee-b}(a) are roughly consistent with the STAR
measurement, there is space to allow for improvements. In the current
work, we have included the smear corrections, which can give an enhancement
in the large $M_{ee}$ region. The higher order corrections beyond
the Born level, such as Sudakov factor \citep{Klein:2018fmp,Klein:2020jom,Li:2019sin,Li:2019yzy,Hatta:2021jcd},
and possible medium effects \citep{Klein:2020jom,Klein:2018fmp,ATLAS:2018pfw,Adam:2018tdm}
may be necessary.


In Fig. \ref{fig:4Angle-P-e}(b), we find that $\left\langle \cos(4\varphi)\right\rangle _{e}$
is sensitive to the nuclear radius. It is possible to extract the
nuclear radius through $\left\langle \cos(4\varphi)\right\rangle _{e}$
in experiments. We also obtain a significant enhancement of $\cos(2\varphi)$
for muon pairs proposed in Ref. \citep{Li:2019sin,Li:2019yzy}.

\begin{acknowledgments}
We would like to thank Wangmei Zha and Zebo Tang for helpful discussion.
This work is partly supported in part by National Natural Science
Foundation of China (NSFC) under Grant No. 11890712, 11890713 (a sub-grant
of 11890710), 12075235, 12135011, by the Strategic Priority Research
Program of Chinese Academy of Sciences under Grant No. XDB34030102
and by National Key Research and Development Program of China with
Grant No. 2018YFE0205200 and 2018YFE0104700.
\end{acknowledgments}

\bibliographystyle{h-physrev}
\phantomsection\addcontentsline{toc}{section}{\refname}\bibliography{UPCref2}

\end{document}